\documentclass[preprint2]{aastex}

\newcommand{\hi}{H{\sc i}}
\newcommand{\hii}{H{\sc ii}}
\newcommand{\ha}{H$\alpha$}
\newcommand{\kms}{km\,s$^{-1}$}
\newcommand{\msol}{$M_{\odot}$}
\newcommand{\eg}{e.\,g.,}
\newcommand{\ie}{i.\,e.,}

\shorttitle{The Environmental Influence on the Evolution of Local Galaxies}
\shortauthors{Bouchard et al.}

\begin{document}
\title{The Environmental Influence on the Evolution of Local Galaxies}

\author{Antoine Bouchard\footnote{Current address: Department of Astronomy, University of Cape Town, Private Bag X3, Rondebosch 7701, Republic of South Africa}}
\affil{Research School of Astronomy and Astrophysics, The Australian National University, Weston ACT 2611, Australia; and Australia Telescope National Facility, Epping, NSW 1710, Australia}
\email{bouchard@ast.uct.ac.za}
\and 
\author{Gary S. Da Costa and Helmut Jerjen}
\affil{Research School of Astronomy and Astrophysics, The Australian National University, Weston ACT 2611, Australia}
\email{gdc@mso.anu.edu.au; jerjen@mso.anu.edu.au}

\begin{abstract}
The results of an H$\alpha$ photometric survey of 30 dwarf galaxies of
various morphologies in the Centaurus A and Sculptor groups are presented.
Of these 30, emission was detected in 13: eight are of late-type, two are
early-type and three are of mixed-morphology. The typical flux detection limit of
$\sim2\times10^{-16}$\,erg s$^{-1}$\,cm$^{-2}$, translates into a Star
Formation Rate (SFR) detection limit of
4$\times10^{-6}\,M_{\odot}\,$yr$^{-1}$.  
In the light of these results, the morphology-density relation is reexamined: It is shown that, despite a number of unaccounted parameters, there are significant correlations between the factors determining the morphological type of a galaxy and its environment. Dwarf galaxies in high density regions have lower current SFR and lower
neutral gas content than their low
density counterparts, confirming earlier results from the Local Group and other denser environments.  
The effect of environment is also seen in the timescale formed from the ratio of blue luminosity to current SFR - dwarfs in higher density environments have larger values, indicating relatively higher past average SFR.
The influence of environment extends very far and no
dwarfs from our sample can be identified as `field' objects. \\


\end{abstract}

\section{Introduction}

Constraining the evolutionary parameters for local galaxies is a difficult
task. Empirical evidence has shown that the physical properties of a stellar system
depend on a poorly constrained mixture of environmental and internal
influences.  For example, galaxies are known to follow a morphology-density
relation where the early-types are found in denser environments than
late-types \citep{dressler1980, postman1984, binggeli1990, whitmore1991}.
This relation also holds for Local Group dwarfs where dwarf Spheroidals
(dSph) and dwarf Ellipticals (dE) are located closer to the Milky Way or
M31 than dwarf Irregulars (dIrr) \citep{einasto1974, vandenbergh1994b}.
However, discrepancies remain within this relation and some galaxies with
similar morphologies, evolving under comparable environmental conditions, can
have different properties, \eg{} the two early-type dwarfs NGC147 and
NGC185 have largely different interstellar medium (ISM) content \citep[see][]{mateo1998}. 
Indeed, it may be that environment plays a subordinate role to local processes 
in the determination of their basic properties such as gas
content and star formation rate \citep{dellenbusch2008}.

The mixed-morphology dwarfs may help to identify these evolutionary
parameters.  Five such objects exist in the Local Group (\ie{} LGS3,
Phoenix, Antlia, DDO210 and Pegasus). While all five contain
sizable amounts of neutral hydrogen (\hi{}) \citep{lo1993, young1997,
st-germain1999, barnes2001} and all show evidence for recent star formation
\citep[][and references therein]{mateo1998}, they also display the smooth
ellipsoidal light distribution typically attributed to the old stellar
population of early-type dwarfs. These intermediate-type galaxies may be evolving objects
currently undergoing the final transition from late to early-type
\citep{mayer2001, pedraz2002, simien2002, grebel2003, derijcke2003,
derijcke2004, vanzee2004, read2005}.

One of the main aspects involved in studying the evolution of a galaxy is the evolution of its stellar population.
As such, it will depend on three parameters: its
past, present and future star formation rates (SFR).  Only the present SFR
can be readily measured, from the intensity of the \ha{} emission
\citep[\eg{}][]{kennicutt1994}. For past and potential future star
formation, this information can be extrapolated from various observational
measurements. For instance, stellar populations
\citep[\eg{}][]{martinez1997} and morphology both reflect past star
formation, and \hi{} measurements allow extrapolation of possible further
star formation.

In this context, the morphology-density relation implies that the past SFR of a galaxy is
regulated by the local environment.  It has been established that the
fraction of galaxies sustaining current star formation decreases near the
projected density center of galaxy clusters \citep{ellingson2001, lewis2002,
gomez2003, balogh2004a, rines2005} and this effect can be measured out to
several virial radii.  One explanation is that ram pressure between the
galaxy's interstellar medium (ISM) and the cluster's intergalactic medium
(IGM) removes the \hi{} gas supply for galaxies dwelling in high density regions, where this phenomenon is most effective due to the large relative velocities.
These \hi{} depleted objects will basically see their on-going star formation terminate due to the lack of gas, 
while the galaxies in the outskirts should not be affected.

For nearby dwarf galaxies, detailed studies of SFR and star formation
history (SFH) \citep{grebel2001, skillman2003a, tolstoy2004}, stellar
population \citep{jerjen2001a, karachentsev2002a, karachentsev2007},
structure \citep{jerjen2000b, jerjen2000c, coleman2004, coleman2005}, \hi{}
content \citep{cote1997, blitz2000, huchtmeier2000, koribalski2004,
chemin2006} and ISM abundances \citep{lee2007} are readily available.  It
is, however, not clear how environmentally driven evolution operates
outside of the high galaxy densities of cluster environments (\eg{} in the
Local Group).  Nevertheless, since many recent efforts have been made to
measure line-of-sight distances to galaxies in the Sculptor and Centaurus A
groups \citep{jerjen1998, jerjen2000a, jerjen2001b, karachentsev2002a,
karachentsev2003, karachentsev2004, karachentsev2007} it is now possible to
use these groups to study the correlation between dwarf evolution
indicators and the direct 3-dimensional galaxy distribution.  The Centaurus
A group is a much denser environment than the Sculptor group and we thereby
expect the former's environmental impact to be greater than the latter's
\citep{bouchard2007}.

This paper presents the results of \ha{} observations of selected dwarf
galaxies in the nearby Centaurus A and Sculptor groups. The aim was to
complete the global picture of current star formation for these galaxies
and the ultimate fate of their ISM.  This enables us to constrain the role
of the environment on the evolution of nearby dwarf galaxies. It also
allows us to probe and explain the local morphology-density relation in a
regime of much lower densities than those of clusters.

\section{Observations}
\subsection{A note on galaxy classification}
The specific galaxy classification scheme used throughout this paper (column 4 of Tables \ref{astrocen} and \ref{astroscl}) comes from the literature and is most often based on integrated photometry from optical B band images \citep[see][]{sandage1984,devaucouleurs1991}. Consequently, these morphologies are a measure of the relative importance and distribution of the young, very bright stars with respect to an older and fainter underlying stellar population. Obviously, such classification is rather loose as it depends on a number of unrelated factors, such as seeing conditions. The classification scheme is refined as new observations become available (\eg{} colour-magnitude diagrams) and galaxies may be swapped from one category to another.

We also use a more general, early vs.\ late type classification, conveying less information about the stellar distribution in the objects. Explicitly, early-type galaxies should be objects lacking any obvious young stellar population, having no ongoing star formation (\ie{} no \ha{} emission) and no detectable ISM (\ie{} no \hi{} emission); late-type galaxies should have an important young stellar population, ongoing star formation and a sizeable amount of ISM. Any other object, \eg{} one dominated by old stars, where \hi{} emission is detected but not \ha{}, should be classified as a mixed-type. In many cases, there has been no attempt to measure \hi{} or \ha{} fluxes and the classification is simply inferred from the Hubble sequence. One of the secondary aims of this paper is to verify the galaxy classification by providing some of the missing \ha{} measurements for early-type dwarf galaxies. 

\subsection{The Centaurus A and Sculptor galaxy groups}
A list of all known galaxies in the Centaurus A (CenA) and Sculptor (Scl) galaxy groups (Tables \ref{astrocen} and \ref{astroscl}, respectively) has been compiled from the catalogues in \citet{cote1997}, \citet{jerjen2000b} and \citet{karachentsev2004, karachentsev2007}. The first reference provides a good overview of the groups with a
bias towards \hi{} rich dIrrs. The latter three articles expand the list of
objects associated with the groups and remove some spurious \hi{}
detections. We believe this combined catalogue to be a representative sample of the real galactic populations residing in the two groups.

\begin{figure}[tbp]
\begin{center}
\includegraphics[width=0.5\textwidth]{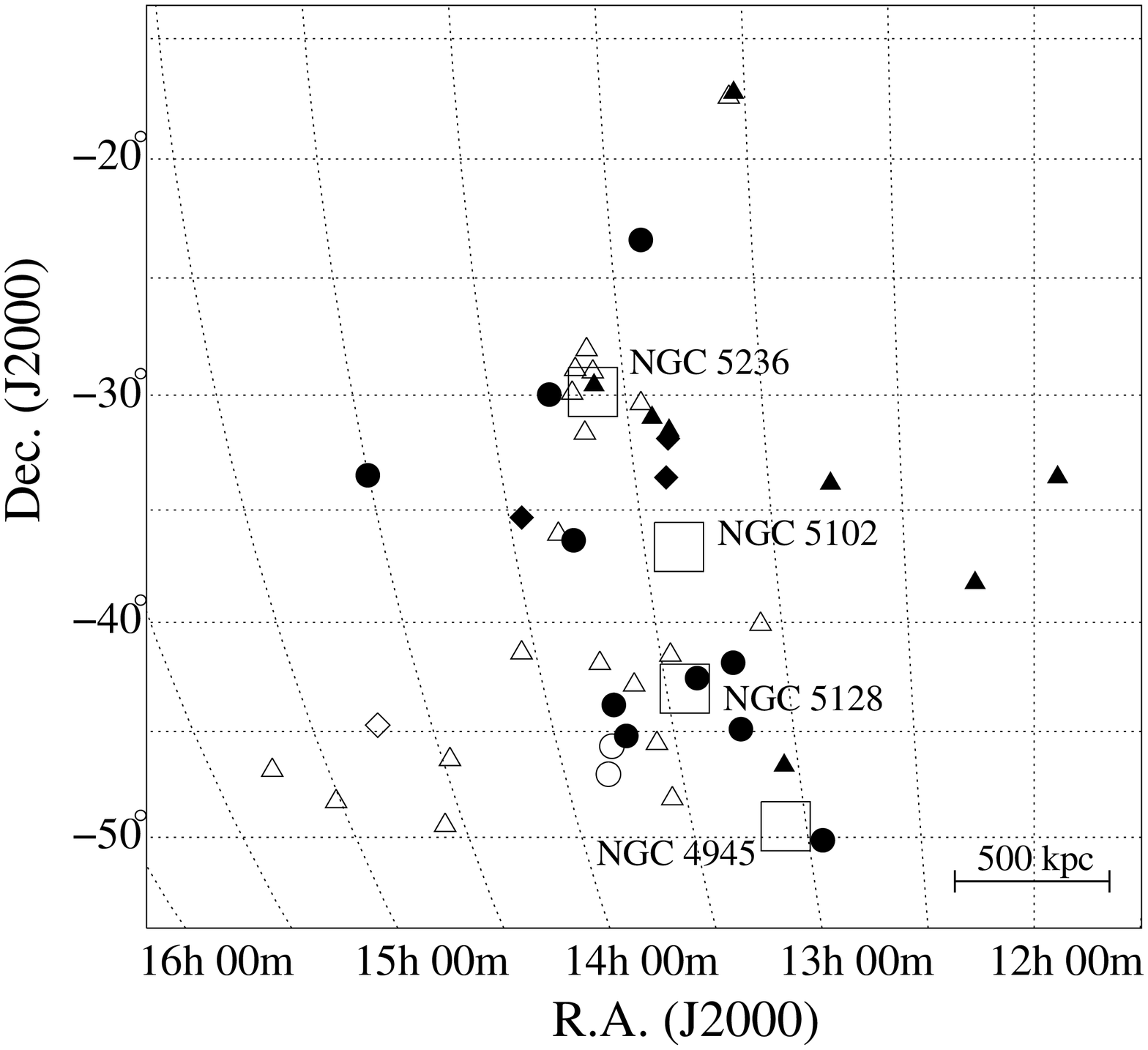}
\includegraphics[width=0.5\textwidth]{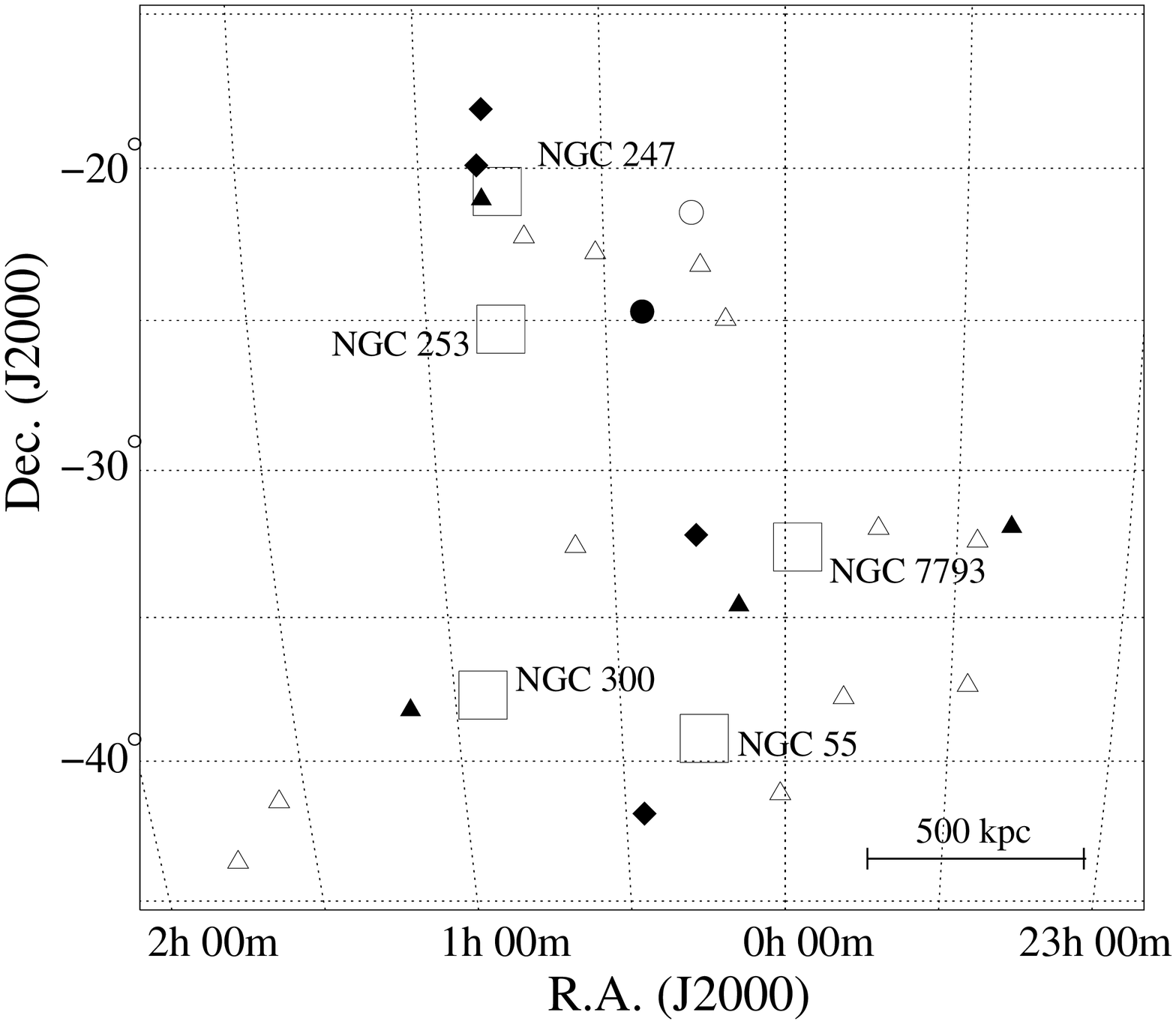}
\caption[Galaxy distribution in CenA and Scl groups]{The galaxy
distribution in CenA (\emph{top}) and Scl (\emph{bottom}). The major
members of the two groups are identified by large squares. The positions of
early-type (\emph{circles}), late-type (\emph{triangles}) and
mixed-morphology dwarfs (\emph{diamonds}) are also marked. Filled symbols
identify the galaxies observed for this paper.}\label{dist}
\end{center}
\end{figure}

Figure \ref{dist} shows the spatial distribution of
galaxies in the two groups and highlights the fundamentally different
environments represented. Not only is the number of early-type dwarfs much
greater in CenA than in Scl, but also the CenA dwarf galaxies are more
clustered around NGC5128 and NGC5236 than the Scl dwarfs are around any of
the major group members.

\subsection{Target selection}

The CenA group (Table \ref{astrocen}) contains 58 galaxies. Of these, 17 are classified as having an early-type morphology, 37 are of late-type and 4 are mixed. The four galaxies of unknown types are presumably of late morphology and have been used as such. The faintest galaxies, [KK2000]55 and LEDA166172, have an apparent integrated magnitude of $m_{\rm B}$\,=\,18.5 and the list is probably complete down to $m_{\rm B}\,\sim\,17$. Based on the entries in Table \ref{astrocen} the CenA group is at an average distance of $D\,=\,4.35$ $\pm$ 0.75\,Mpc (standard deviation), and has an average systemic heliocentric velocity of $V_{\odot}\,=\,408$ $\pm$ 275\,\kms{}. 

The Scl group (Table \ref{astroscl}) is composed of 27 galaxies, 1 of early type, 22 late types and 4 mixed. These are at an average distance of $D\,=\,3.95$ $\pm$ 1.6\,Mpc and average velocity of $V_{\odot}\,=\,290$ $\pm$ 190\,\kms{}, based on Table \ref{astroscl} data. The faintest object has $m_{\rm B}\,=\,18$ and the group is probably complete to $m_{\rm B}\,\sim\,16$. 

The focus of our observations was
to constrain the characteristics of galaxies at the low end of the
luminosity function because these objects are more likely to be affected by
environmental parameters than the ones with higher luminosities.
Consequently, little attention was given to brighter members of the groups.
We observed a selection of dwarf galaxies within the two groups, provided they are of low luminosity ($-10\,\gtrsim\,M_{\rm B}\,\gtrsim\,-14$) and no \ha{} measurements could be found in the literature. A consequence of these criteria is that nearly all known low mass early-type galaxies of the CenA and Scl groups were observed, randomly leaving out some of the faintest ones ($M_{\rm B}\,>\,-10.5$) because of limited observing time. Moreover, some objects (mostly late-types) satisfying the luminosity constraint were already observed in \ha{} by \citet{cote2008} hence were also left out of the present study.
For comparison purposes, previously observed objects were randomly added to our selection. Our final sample can be found in Table \ref{objects} and was highlighted in Tables \ref{astrocen}, \ref{astroscl} with asterisks in front of the galaxy name, as well as in Figure \ref{dist} with filled symbols.

\begin{table*}[tbp]
\begin{tiny}
\begin{center}
\caption{Astrometric and photometric properties for CenA group galaxies}\label{astrocen}
\begin{tabular}{l c c c c c c c}
\hline
\hline
Galaxy  &  RA      & Dec     & Type & $m_{\rm B}$ & $D$ & $V_{\odot}$ & Ref.\\
	& (J2000) & (J2000) &      &                &   (Mpc)  & (km\,s$^{-1}$) &     \\
\hline
{}$\star{}$ ESO379-G007 &  11:54:43 &  -33:33:36 &  dIrr &  16.6$\pm$0.09 &  5.2$\pm$0.5&  641$\pm$4 &  1, 2\\
{}$\star{}$ ESO321-G014 &  12:13:49 &  -38:13:53 &  IBm &  15.21$\pm$0.09 &  3.2$\pm$0.3&  610$\pm$3 &  1, 2\\
{}ESO381-G018 & 12:44:42 & -35:58:00 & Irr & 15.72$\pm$0.1 & 5.3$\pm$0.4& \nodata & 3\\
{}$\star{}$ ESO381-G020 &  12:46:00 &  -33:50:13 &  IB(s)m &  14.24$\pm$0.09 &  5.5$\pm$0.5&  589$\pm$2 &  2, 3\\
{}ESO443-G009 & 12:54:53 & -28:20:27 & dIm & 17.06$\pm$0.1 & 5.8$\pm$0.5& \nodata & 3\\
{}$\star{}$ ESO219-G010 &  12:56:09 &  -50:08:38 &  dE,N &  16.4$\pm$0.2 &  4.8$\pm$0.4& \nodata &  4, 5\\
{}$\star{}$ UGCA319 &  13:02:14 &  -17:14:15 &  IB(s)m &  15.33$\pm$0.1 &  \nodata&  755$\pm$4&  2, 6\\
{}DDO161 & 13:03:16 & -17:25:23 & IB(s)m\_sp & 13.5$\pm$0.1 & \nodata& $742\pm2$& 2, 6\\
{}$\star{}$ ESO269-G037 &  13:03:33 &  -46:35:06 &  dSph &  16.29$\pm$0.09 &  3.5$\pm$0.4&  744$\pm$2  &  1, 7\\
{}[CFC97]Cen6 & 13:05:02 & -40:04:58 & dIrr & 16.33$\pm$0.1 & 5.8$\pm$0.5& $614\pm1$& 3, 7\\
{}NGC4945 & 13:05:27 & -49:28:05 & SB(s)cd(Sy2) & 9.4$\pm$0.2 & 3.8$\pm$0.3& $563\pm3$& 2, 3\\
{}ESO269-G058 & 13:10:32 & -46:59:27 & I0 pec & 13.3$\pm$0.1 & 3.8$\pm$0.3& \nodata  & 3\\
{}$\star{}$ CenA-dE1 &  13:12:45 &  -41:49:57 &  dSph &  17.75$\pm$0.1 &  4.2$\pm$0.3& \nodata &  3, 5\\
{}$\star{}$ ESO269-G066 &  13:13:09 &  -44:53:24 &  dE,N &  14.59$\pm$0.08 &  4.1$\pm$0.5&  780$\pm$30 &  4, 5\\
{}$\star{}$ HIPASSJ1321-31 &  13:21:08 &  -31:31:45 &  dIrr &  17.1$\pm$0.1 &  5.2$\pm$0.3&  571$\pm$3 &  2, 6\\
{}$\star{}$ CenA-dE2 &  13:21:33 &  -31:52:43 &  dE/dIm &  18.1$\pm$0.2 & \nodata& \nodata &  5\\
{}[KK98]196 & 13:21:47 & -45:03:48 & IBm & 16.1$\pm$0.1 & 4.0$\pm$0.3& \nodata & 3\\
{}NGC5102 & 13:21:57 & -36:37:48 & SB(s)b & 10.0$\pm$0.2 & 3.4$\pm$0.4& $468\pm2$& 1, 2\\
{}$\star{}$ SGC1319.1-4216 &  13:22:02 &  -42:32:07 &  dE &  15.7$\pm$0.1 &  3.9$\pm$0.3& \nodata &  3, 5\\
{}[KK2000]55 & 13:22:12 & -42:43:51 & dSph & 18.5$\pm$0.1 & 3.9$\pm$0.3& \nodata & 3\\
{}$\star{}$ [CFC97]Cen8 &  13:22:56 &  -33:34:22 &  Im/dE &  17.65$\pm$0.08 & \nodata& \nodata &  5\\
{}$\star{}$ AM1320-230 &  13:23:29 &  -23:23:35 &  dE &  17.53$\pm$0.08 & \nodata& \nodata &  5\\
{}$\star{}$ AM1321-304 &  13:24:36 &  -30:58:19 &  dIm &  16.7$\pm$0.1 &  4.6$\pm$0.5&  485$\pm$3&  1\\
{}NGC5128 & 13:25:27 & -43:01:08 & S0pec(sy2) & 7.84$\pm$0.09 & 3.8$\pm$0.4& $556\pm10 $& 2, 8\\
{}IC4247 & 13:26:44 & -30:21:44 & S? & 14.41$\pm$0.09 & 5.0$\pm$0.4& $415\pm30 $& 3\\
{}ESO324-G024 & 13:27:37 & -41:28:50 & IABm & 12.91$\pm$0.09 & 3.7$\pm$0.4& $516\pm3 $& 1, 2\\
{}NGC5206 & 13:33:44 & -48:09:04 & SB0 & 11.62$\pm$0.09 & 3.5$\pm$0.3& $571\pm10 $& 3, 6\\
{}ESO270-G017 & 13:34:47 & -45:32:51 & SB(s)m: & 11.69$\pm$0.09 & 4.3$\pm$0.8& $828\pm2 $& 2, 3\\
{}UGCA365 & 13:36:31 & -29:00:00 & Im & 15.49$\pm$0.09 & 5.2$\pm$0.4& $571\pm1  $& 3, 7\\
{}$\star{}$ [KK98]208 &  13:36:35 &  -29:34:17 &  dIrr &  14.3$\pm$0.1 &  4.7$\pm$0.4& \nodata &  1\\
{}NGC5236 & 13:37:00 & -29:51:56 & SAB(s)c & 8.22$\pm$0.09 & 5.2$\pm$0.4& $513\pm2$& 2, 3\\
{}DEEPJ1337-33 & 13:37:00 & -33:21:47 & ? & 17.3$\pm$0.1 & 4.5$\pm$0.5& \nodata & 6\\
{}ESO444-G084 & 13:37:20 & -28:02:42 & Im & 15.01$\pm$0.09 & 4.6$\pm$0.5& $587\pm3$& 1, 2\\
{}HIPASSJ1337-39 & 13:37:25 & -39:53:52 & Im & 16.5$\pm$0.1 & 4.9$\pm$0.5& $492\pm4$ & 2, 6\\
{}NGC5237 & 13:37:39 & -42:50:49 & I0? & 13.26$\pm$0.09 & 3.4$\pm$0.3& $361\pm4$& 2, 3\\
{}NGC5253 & 13:39:55 & -31:38:24 & Im\_pec & 11.17$\pm$0.09 & 3.7$\pm$0.3& $407\pm3$& 2, 9\\
{}IC4316 & 13:40:18 & -28:53:38 & IBm? pec & 15.0$\pm$0.2 & 4.4$\pm$0.4& $670\pm50$& 1\\
{}NGC5264 & 13:41:36 & -29:54:47 & IB(s)m & 12.6$\pm$0.2 & 4.5$\pm$0.5& $478\pm3$& 1, 2\\
{}[KK2000]57 & 13:41:38 & -42:34:55 & dSph & 18.1$\pm$0.1 & 3.9$\pm$0.3& \nodata & 3\\
{}$\star{}$ AM1339-445 &  13:42:05 &  -45:12:18 &  dE &  16.32$\pm$0.1 &  3.5$\pm$0.3& \nodata &  5, 10\\
{}$\star{}$ LEDA166172 &  13:43:36 &  -43:46:11 &  dSph &  18.5$\pm$0.1 &  3.6$\pm$0.4& \nodata &  1\\
{}ESO325-G011 & 13:45:00 & -41:51:40 & IB(s)m & 14.02$\pm$0.09 & 3.4$\pm$0.4& $545\pm2$& 1, 2\\
{}$\star{}$ CenA-dE3 &  13:46:00 &  -36:20:15 &  dE &  17.4$\pm$0.2 & \nodata& \nodata &  5\\
{}AM1343-452 & 13:46:16 & -45:41:05 & dSph & 17.6$\pm$0.1 & 3.7$\pm$0.3& \nodata & 5, 10\\
{}$\star{}$ CenA-dE4 &  13:46:40 &  -29:58:41 &  dE &  17.6$\pm$0.1 & \nodata& \nodata &  5\\
{}CenN & 13:48:09 & -47:33:54 & ? & 17.5$\pm$0.6 & 3.7$\pm$0.3& \nodata & 3\\
{}HIPASSJ1348-37 & 13:48:33 & -37:58:03 & ? & 16.9$\pm$0.1 & 5.8$\pm$0.5& \nodata & 3\\
{}LEDA166179 & 13:48:46 & -46:59:46 & dSph? & 18$\pm$0.1 & 4.0$\pm$0.4& \nodata & 1\\
{}ESO383-G087 & 13:49:18 & -36:03:41 & SB(s)m & 11$\pm$0.09 & 3.5$\pm$0.3& $326\pm2$& 2, 3\\
{}HIPASSJ1351-47 & 13:51:22 & 47:00:00 & ? & 17.5$\pm$0.1 & 5.7$\pm$0.5& \nodata & 3\\
{}$\star{}$ ESO384-G016 &  13:57:01 &  -35:20:01 &  dS0/Im &  15.11$\pm$0.06 &  4.5$\pm$0.4&  560$\pm$30&  4, 5\\
{}NGC5408 & 14:03:20 & -41:22:39 & IB(s)m & 12.2$\pm$0.2 & 4.8$\pm$0.5& $506\pm3$& 1, 2\\
{}UKS1424-460 & 14:28:03 & -46:18:06 & IB(s)m & 16.5$\pm$0.1 & 3.6$\pm$0.4& $390\pm2$& 1, 2\\
{}$\star{}$ CenA-dE5 &  14:30:05 &  -33:28:45 &  dE &  18.4$\pm$0.1 & \nodata& \nodata &  5\\
{}ESO222-G010 & 14:35:02 & -49:25:14 & dIrr & 16.3$\pm$0.1 & \nodata& $622\pm5$& 2\\
{}ESO272-G025 & 14:43:25 & -44:42:18 & dE/dIrr & 14.8$\pm$0.1 & \nodata& $629\pm1$& 7\\
{}ESO223-G009 & 15:01:08 & -48:17:25 & IAB(s)m & 14.8$\pm$0.1 & 6.4$\pm$0.5& $588\pm2$& 2, 3\\
{}ESO274-G001 & 15:14:13 & -46:48:33 & Sd & 12.0$\pm$0.2 & 3.1$\pm$0.3& $522\pm2 $& 2, 3\\
\hline
\end{tabular}
\begin{itemize}
\item {\sc Note} --- Galaxies with the $\star$ symbol preceding there name, are objects that were observed in this study. Units of right ascension are hours, minutes, and seconds, and units of declination are degrees, arcminutes, and arcseconds (J2000.0).
\item {\sc References} --- (1) \citet{karachentsev2002a}; (2)
\citet{koribalski2004}; (3) \citet{karachentsev2007}; (4)
\citet{jerjen2000a}; (5) \citet{jerjen2000b}; (6) \citet{karachentsev2004};
(7) \citet{bouchard2006};
(8) \citet{israel1998}; (9) \citet{sakai2004}; (10) \citet{rejkuba2006}
\end{itemize}
\end{center}
\end{tiny}
\end{table*}

\begin{table*}[tbp]
\begin{center}
\begin{tiny}
\begin{center}
\caption{Astrometric and photometric properties for Sculptor group galaxies}\label{astroscl}
\begin{tabular}{l c c c c c c c}
\hline
\hline
Galaxy & RA      & Dec     & Type & $m_{\rm B}$ & $D$ & $V_{\odot}$ & Ref.\\
       & (J2000) & (J2000) &      &                &    (Mpc)  & (km\,s$^{-1}$)     \\
\hline
{}$\star{}$ WHIB2317-32 & 23:20:37 & -31:54:21 & dIrr & 18$\pm$0.1 & \nodata & $68$& 1, 2\\
{}UGCA438 & 23:26:27 & -32:23:26 & IB(s)m pec: & 13.9$\pm$0.1 & 2.2$\pm$0.2 & $62$& 1, 3\\
{}ESO347-G017 & 23:26:57 & -37:20:47 & SB(s)m:sp & 14.7$\pm$0.1 & 7.0$\pm$0.4 & $692\pm4$ & 4, 5\\
{}UGCA442 & 23:43:46 & -31:57:24 & SB(s)m: sp & 13.6$\pm$0.1 & 4.3$\pm$0.5 & $267\pm2$ & 1, 5\\
{}ESO348-G009 & 23:49:23 & -37:46:19 & IBm & 13.6$\pm$0.1 & 6.5$\pm$0.4 & $648\pm4$ & 4, 5\\
{}NGC7793 & 23:57:50 & -32:35:28 & SA(s)d & 9.72$\pm$0.09 & 3.9$\pm$0.4 & $227\pm2$ & 1, 5\\
{}[CFC97]SC18 & 00:00:56 & -41:08:53 & dIrr & 17.5$\pm$0.1 & 2.1$\pm$0.2 & $151$ & 1, 2, 4\\
{}$\star{}$ ESO349-G031 & 00:08:13 & -34:34:42 & IBm & 15.5$\pm$0.1 & 3.2$\pm$0.3 & $221\pm6$ & 1, 2, 6\\
{}NGC24 & 00:09:57 & -24:57:47 & SA(s)c & 12.4$\pm$0.1 & \nodata & $554\pm2$ & 5, 7\\
{}NGC45 & 00:14:04 & -23:10:55 & SA(s)dm & 11.5$\pm$0.1 & \nodata & $467\pm2$ & 5, 7\\
{}NGC55 & 00:14:54 & -39:11:48 & SB(s)m:sp & 8.8$\pm$0.1 & 1.9$\pm$0.1 & $129\pm2$ & 8, 5\\
{}NGC59 & 00:15:25 & -21:26:38 & dS0 & 12.97$\pm$0.03 & 4.4$\pm$0.4 & $367$ & 2, 9, 10\\
{}$\star{}$ ESO410-G005 & 00:15:31 & -32:10:55 & dE/Im & 14.8$\pm$0.1 & 1.9$\pm$0.2 & $159\pm2$ & 1, 11\\
{}$\star{}$ Scl-dE1 & 00:23:51 & -24:42:18 & dE & 17.7$\pm$0.2 & 4.2$\pm$0.4 &\nodata & 1, 9\\
{}$\star{}$ ESO294-G010 & 00:26:33 & -41:51:19 & dS0/Im & 15.53$\pm$0.04 & 1.71$\pm$0.07 & $107\pm1$ & 9, 10, 11\\
{}ESO473-G024 & 00:31:23 & -22:45:58 & Im & 16.0$\pm$0.1 & 5.9$\pm$0.4 & $540\pm4$ & 4, 5\\
{}[CFC97]SC24 & 00:36:39 & -32:34:25 & dIrr & 18.0$\pm$0.1 & 1.7$\pm$0.2 & $79$ & 1, 2, 4\\
{}IC1574 & 00:43:04 & -22:14:49 & IB(s)m & 14.4$\pm$0.1 & 4.9$\pm$0.6 & $363\pm4$ & 1, 5\\
{}NGC247 & 00:47:09 & -20:45:37 & SAB(s)d & 9.9$\pm$0.1 & 3.7$\pm$0.4 & $156\pm2$ & 1, 5, 6\\
{}NGC253 & 00:47:33 & -25:17:18 & SAB(s)c & 7.9$\pm$0.1 & 3.9$\pm$0.4 & $243\pm2$ & 1, 5\\
{}$\star{}$ ESO540-G030 & 00:49:21 & -18:04:34 & dE/Im & 16.37$\pm$0.07 & 3.2$\pm$0.1 & $224\pm3$ & 9, 10, 11\\
{}$\star{}$ DDO6 & 00:49:49 & -21:00:54 & IB(s)m: & 15.2$\pm$0.1 & 3.3$\pm$0.2 & $294\pm4$ & 1, 5\\
{}$\star{}$ ESO540-G032 & 00:50:25 & -19:54:23 & dE/Im & 16.44$\pm$0.08 & 3.7$\pm$0.2 & $228\pm1$ & 12, 11\\
{}NGC300 & 00:54:54 & -37:41:04 & SA(s)d & 9.0$\pm$0.1 & 2.2$\pm$0.2 & $146\pm2$ & 1, 5\\
{}$\star{}$ AM0106-382 & 01:08:22 & -38:12:33 & dIm & 16.3$\pm$0.1 & 6.1$\pm$0.4 & $645$ & 2, 4\\
{}NGC625 & 01:35:06 & -41:26:05 & SB(s)m?sp & 11.6$\pm$0.1 & 4.1$\pm$0.4 & $396\pm3$ & 1, 5\\
{}ESO245-G005 & 01:45:04 & -43:35:53 & IB(s)m & 12.7$\pm$0.1 & 4.4$\pm$0.5 & $391\pm2$ & 1, 5\\
\hline
\end{tabular}
\begin{itemize}
\item {\sc Note} --- Galaxies with the $\star$ symbol preceding there name, are objects that were observed in this study. Units of right ascension are hours, minutes, and seconds, and units of declination are degrees, arcminutes, and arcseconds (J2000.0).
\item {\sc References} --- (1) \citet{karachentsev2004}; (2) \citet{cote1997}; (3) \citet{karachentsev2002a}; (4) \citet{skillman2003a}; (5) \citet{koribalski2004}; (6) \citet{karachentsev2006}; (7) \citet{devaucouleurs1991}; (8) \citet{pietrzynski2006}; (9) \citet{jerjen2000b}; (10) \citet{jerjen1998}; (11) \citet{bouchard2005}; (12) \citet{dacosta2007}
\end{itemize}
\end{center}
\end{tiny}
\end{center}
\end{table*}

\subsection{\ha{} observations}
The \ha{} line emission was imaged using the Australian National University
2.3\,m Telescope at the Siding Spring Observatory. The Nasmyth Imager,
equipped with the SITe 1024$\times$1024 thinned CCD, has a circular field
of view of 6$^{\prime}.$6 in diameter and pixel size of
0$^{\prime\prime}.$59.  For the Scl galaxies, three exposures of 900
seconds were taken with a narrowband 7 nm wide ``on-band'' filter centered
on a wavelength of 657.5 nm, encompassing the \ha{} rest frequency of 656.3
nm. Three 300 seconds exposures of the same field, using an 8 nm wide \ha{}
``off-band'' filter centered on 645 nm, were used to estimate the \ha{}
continuum.  For the galaxies in CenA, bad weather conditions permitted only
one 900 seconds on-band and one 300 seconds off-band exposure per galaxy.
Additionally, several Planetary Nebul\ae{} (PNs) from the list by
\citet{dopita1997} were observed for flux calibration purposes.

After subtraction of the bias, flat fielding, removal of the sky background
and combination of the multiple images to increase the signal-to-noise
ratio (where multiple images were available), the intensity of the off-band
images were scaled to match that of their counterpart on-band images by
comparing the intensities of foreground stars.  This normalisation removed
the effects caused by the differences in the bandpass of each filter as
well as the different integration times.  The off-band images were then
subtracted from the on-band, thus leaving only the line emission in the
resulting frames.

The \ha{} line total flux, \ie{} the sum of the flux value of each pixel
where a significant amount of emission is detected, was compared with the
data from the PNs. In practice, the sum is done on all pixels situated
inside the 3$\sigma$ contour level, where $\sigma$ is taken to be the
standard deviation of the noise distribution in a sky region of final
image. This value includes all noise sources such as readout and photon
noise.  The flux was corrected to take into account the difference in
transmission caused by the varying redshifts at which the \ha{} line is observed and the shape of the bandpass of the on-band \ha{} filter.
When the redshift of the target galaxy was unknown, the \ha{} transmission
was assumed to be the most common value for other galaxies of the same
group.

\subsection{\ha{} results}
\begin{figure*}[tbp]
\includegraphics[width=0.95\textwidth]{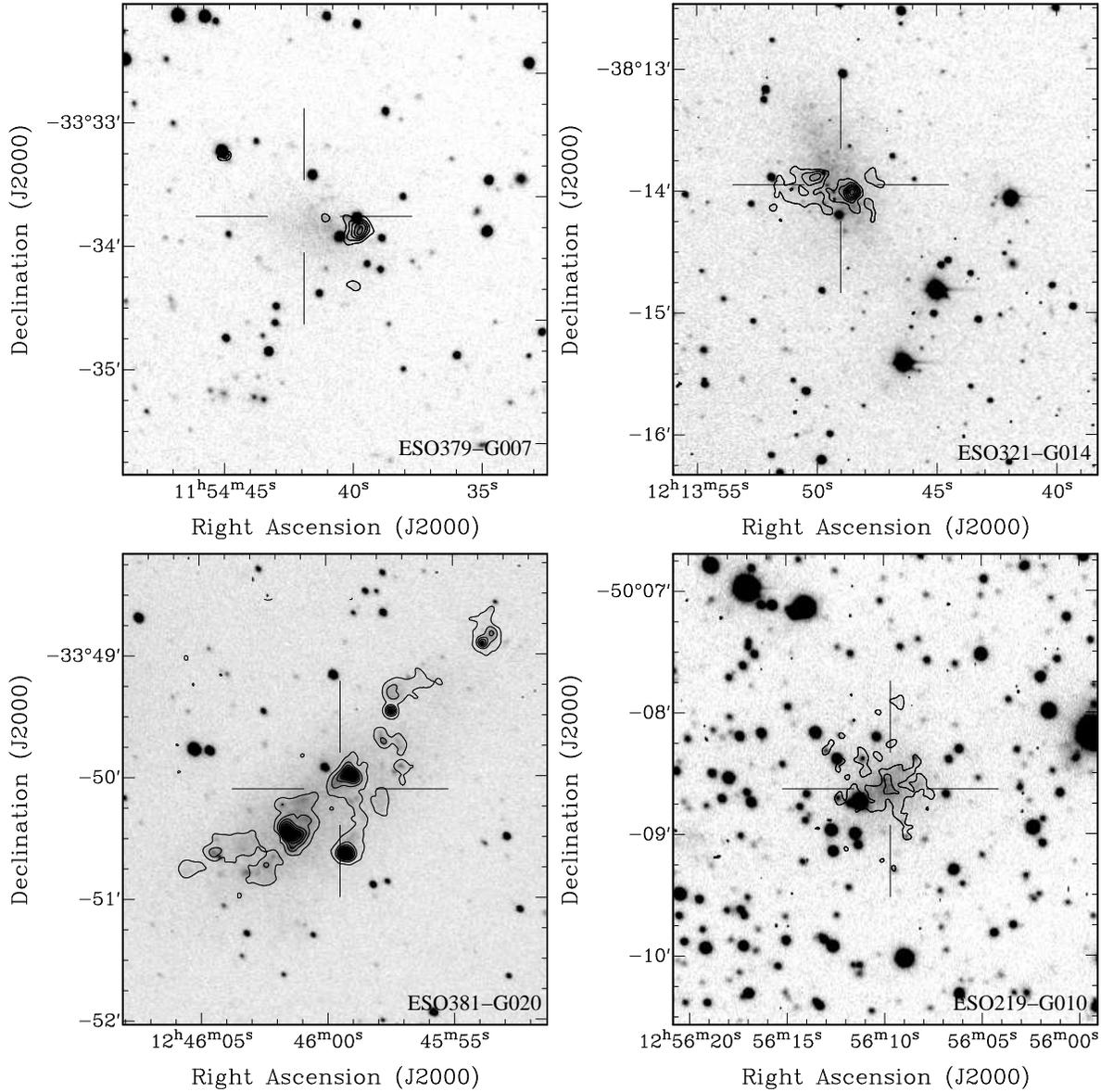}
\caption[\ha{} detection of ESO379-G007, ESO321-G014, ESO381-G020 and
ESO219-G010]{\ha{} contours overlaid on the \ha{} continuum and line image
of the four CenA group members ESO379-G007 (\emph{top-left}), ESO321-G014
(\emph{top-right}), ESO381-G020 (\emph{bottom-left}) and ESO219-G010
(\emph{bottom-right}).  The first contour level represent the 3$\sigma$
significance level. Each following contour is a further significance
increase of 3$\sigma$ except for ESO381-G020, where the increase is
12$\sigma$. The
1$\sigma$ level corresponds to a flux density of $\sim5\times10^{-18}$ erg
s$^{-1}$ cm$^{-2}$ arcsec$^{-2}$.
The images are approximately 4\arcmin$\times$4\arcmin{} in size. The crosshairs indicate the optical center of the galaxies.}\label{ff1}
\end{figure*}

\begin{figure*}[tbp]
\includegraphics[width=0.95\textwidth]{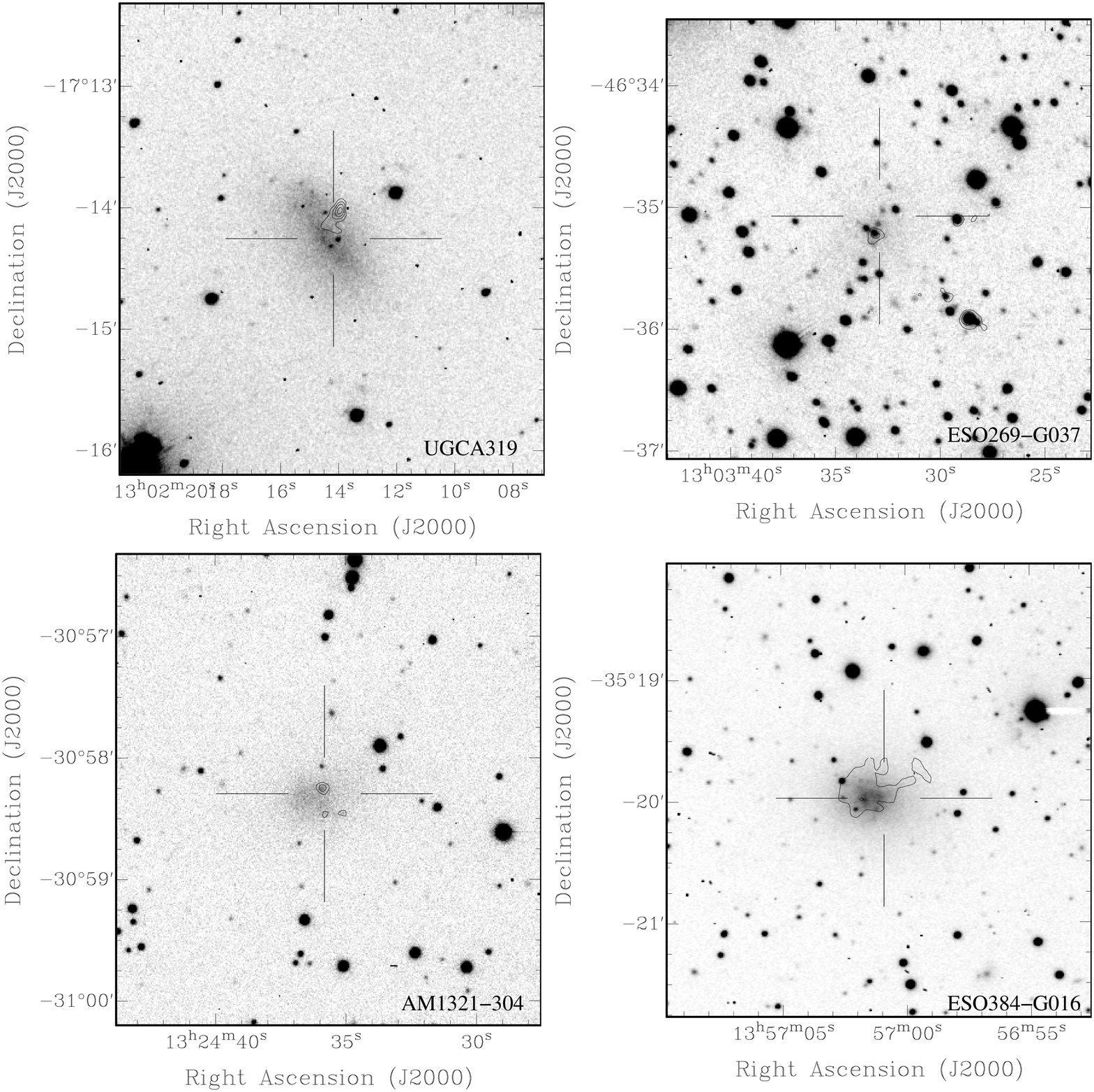}
\caption[\ha{} detection of UGCA319, ESO269-G037, AM1321-304 and
ESO384-G016]{\ha{} contours overlaid on the \ha{} continuum and
line image of the four CenA group members UGCA319 (\emph{top-left}), ESO269-G037 (\emph{top-right}), AM1321-304
(\emph{bottom-left}) and ESO384-G016 (\emph{bottom-right}). 
The first contour level represent the 3$\sigma$ significance level and each
additional contour represents a further increase of 3$\sigma$. The
1$\sigma$ level corresponds to a flux density of $\sim5\times10^{-18}$ erg
s$^{-1}$ cm$^{-2}$ arcsec$^{-2}$.
The images are approximately 4\arcmin$\times$4\arcmin{} in size. The crosshairs indicate the optical center of the galaxies.}\label{ff2}
\end{figure*}

\begin{figure*}[tbp]
\includegraphics[width=0.95\textwidth]{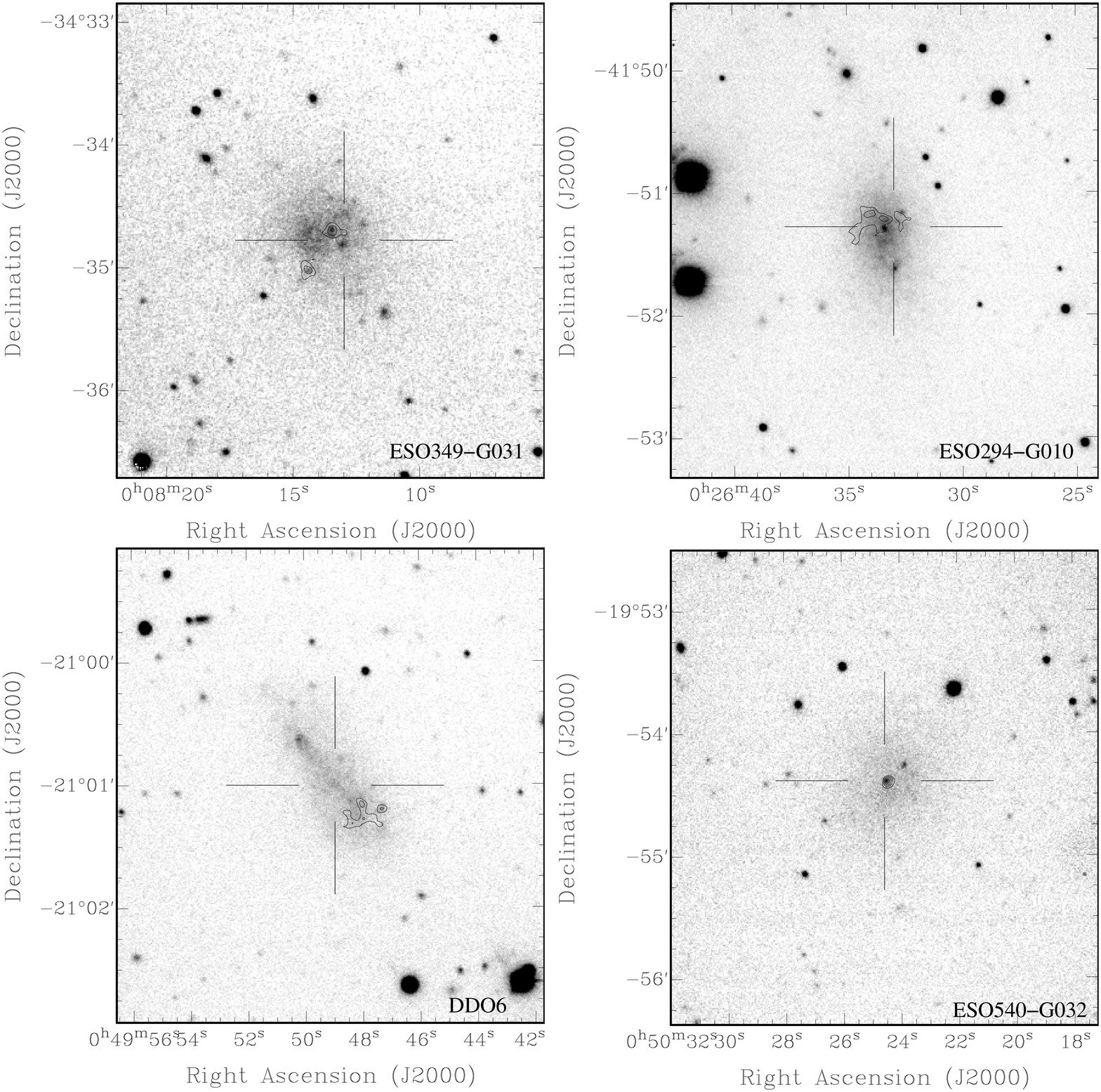}
\caption[\ha{} detection of ESO349-G031, ESO294-G010, DDO6 and
ESO540-G032]{\ha{} contours overlaid on the \ha{} continuum and line image
of the four Scl group members ESO349-G031 (\emph{top-left}), ESO294-G010
(\emph{top-right}), DDO6 (\emph{bottom-left}) and ESO540-G032
(\emph{bottom-right}). 
The first contour level represents the 3$\sigma$ significance level and
each additional contour represents a further increase of 3$\sigma$. The
1$\sigma$ level corresponds to a flux density of $\sim3\times10^{-18}$ erg
s$^{-1}$ cm$^{-2}$ arcsec$^{-2}$.  The images are approximately
4\arcmin$\times$4\arcmin{} in size. The crosshairs indicate the optical center of the galaxies.}\label{ff3}
\end{figure*}

\begin{figure}[tbp]
\begin{center}
\includegraphics[width=0.45\textwidth]{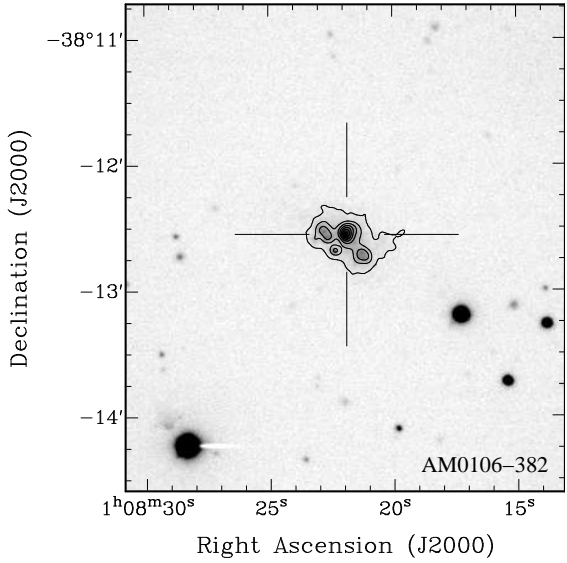}
\caption[\ha{} detection of AM0106-382]{\ha{} contours overlaid on the
\ha{} continuum and line image
the Scl group member AM0106-382.
The first contour level represents the 3$\sigma$ significance level and
each additional contour represents a further increase of 12$\sigma$.  The
1$\sigma$ level corresponds to a flux density of $\sim3\times10^{-18}$ erg
s$^{-1}$ cm$^{-2}$ arcsec$^{-2}$.  The image is approximatively
4\arcmin$\times$4\arcmin{}. The crosshair indicates the optical center of the galaxy.}\label{ff4}
\end{center}
\end{figure}

Figure \ref{ff1} shows the location of the \ha{} line emission in the dIrr
ESO379-G007, the IBm ESO321-G014, the IB(s)m ESO381-G020 and the dE,N
ESO219-G010. Similarly, Figure \ref{ff2} shows the \ha{} line emission in
the IABm ESO269-G037, the dIm AM1321-304, the dS0/Im ESO384-G016 and the
IB(s)m UGCA319.  All eight galaxies are members of CenA.  The galaxies of
Scl are presented in Figure \ref{ff3}, for the IBm ESO349-G031, the dS0/Im
ESO294-G010, the IB(s)m DDO6 and the dE/Im ESO540-G032 while the dIm galaxy
AM0106-382 is in Figure \ref{ff4}.

The total integrated \ha{} line flux for these objects are found in Table
\ref{objects}.  The table also lists several non-detections where an
upper-limit on the \ha{} flux is listed. This limit is taken to be
approximately 3 times the typical noise over an area of 100 pixels square
or $6\arcsec{}\times6\arcsec{}$.

\begin{table}[tbp]
\begin{center}
\caption{\ha{} flux table, all observed objects}\label{objects}
\begin{tabular}{l c}
\hline
\hline
Galaxy		& \ha{} line flux\\
		& (10$^{-15}$ erg s$^{-1}$ cm$^{-2}$)\\
\hline
\multicolumn{2}{c}{\emph{CenA group}}\\
\hline
ESO379-G007	& 7.3$\pm$1.9	\\
ESO321-G014	& 67$\pm$18	\\
ESO381-G020	& 228$\pm$61	\\
ESO219-G010	& 27.3$\pm$7.8$^\dagger$	\\
UGCA319		& 7.2$\pm$1.9	\\
ESO269-G037	& 3.2$\pm$0.9$^\dagger$	\\
CenA-dE1	& $<0.25$	\\
ESO269-G066	& $<0.25$	\\
HIPASS1321-31	& $<0.25$	\\
CenA-dE2	& $<0.25$	\\
SGC1319.1-4216	& $<0.25$	\\
{}[CFC97]Cen8	& $<0.25$	\\
AM1320-230	& $<0.25$	\\
AM1321-304	& 2.2$\pm$0.6	\\
{}[KK98]208	& $<0.25$	\\
AM1339-445	& $<0.25$	\\
LEDA166172	& $<0.25$	\\
CenA-dE3	& $<0.25$	\\
CenA-dE4	& $<0.25$	\\
ESO384-G016	& 13.6$\pm$3.7	\\
CenA-dE5	& $<0.25$	\\
\hline
\multicolumn{2}{c}{\emph{Scl group}}\\
\hline
WHIB2317-32	& $<0.15$	\\
ESO349-G031	& 5.9$\pm$1.6	\\
ESO410-G005	& $<0.15$	\\
Scl-dE1		& $<0.15$	\\
ESO294-G010	& 13.3$\pm$3.6	\\
ESO540-G030	& $<0.15$	\\
DDO 6		& 6.2$\pm$1.7	\\
ESO540-G032	& 0.8$\pm$0.3	\\
AM0106-382	& 71$\pm$19	\\
\hline
\end{tabular}
\begin{footnotesize}
\begin{itemize}
\item {\sc Note} --- The values marked with $^\dagger$ may be false detections, see text for details.
\end{itemize}
\end{footnotesize}
\end{center}
\end{table}

\subsection{Comments on individual objects of the CenA group}
\subsubsection{ESO379-G007}
An \hii{} region in this dIrr galaxy is situated $\sim10$\arcsec{} (or
250\,pc projected distance at $D = 5.2$\,Mpc) west of the optical center
and is slightly extended. Concordantly, the colour-magnitude diagram (CMD)
for this object also reveals some young main sequence stars
\citep{karachentsev2002a}.  The arc-shape of the \ha{} emission and its
position with respect to the optical center may suggest that ram pressure
is operating. This however seems unlikely as this galaxy is one of the most
isolated objects of the CenA group, situated at $1.3\pm0.1$\,Mpc (3D distance) from
NGC5236 ($2\pm0.8$\,Mpc from NGC5102).

\subsubsection{ESO321-G014}
This IABm galaxy has an \hii{} point source that is displaced from the
center of the galaxy by about 15\arcsec{} or 230\,pc projected distance, at
$D = 3.2$\,Mpc. It is accompanied by low level diffuse emission that
extends along the minor axis of the galaxy. The CMD shows
young main sequence and some upper-AGB stars --- signature of the presence
of an intermediate age population, \ie{} 1\,Gyr and older --- are found
\citep{karachentsev2002a}.  This galaxy is a fairly isolated object of the
CenA group, situated at $640\pm150$\,kpc (3D) from NGC5102.

\subsubsection{ESO381-G020}
This dwarf clearly contains both upper-AGB and young main sequence stars
\citep{karachentsev2007}. It has several \ha{} point sources and a
considerable amount of diffuse emission. As for the two preceding galaxies,
it is a relatively isolated object at $780\pm200$\,kpc (3D) from NGC5236 ($2.1\pm0.8$\,Mpc
from NGC5102).

\subsubsection{ESO219-G010}
This galaxy seems to have faint diffuse \hii{} emission that is not accompanied by
any point source.  However, previous investigations failed to detect
\hi{} for this object \citep{beaulieu2006, bouchard2007}. The upper-limits,
$M_{\rm HI} < 7\times10^5\,M_{\odot}$ and $M_{\rm HI}/L_{\rm B} <
0.03\,M_{\odot}\,L_{\odot}^{-1}$ are very low and the \hi{} properties are
consistent with that of other early-type dwarfs.  The \ha{} detection could be interpreted
as a spurious detection.  Careful inspection of the \ha{}, off
band and residual images reveal no signs of an under-subtracted continuum but different seeing conditions in the off-band and on-band images is likely to have altered our capacity of making a good continuum estimate. 
No CMD is available for this galaxy to help resolve this issue, additional data is required to confirm this detection. It is situated at $1\pm0.5\,$Mpc (3D) from
NGC4945.

\subsubsection{UGCA319}
This IB(s)m galaxy contains an \ha{} point source  $\sim15$\arcsec{} or
320\,pc projected distance (at $D = 4.4$ Mpc) north of its center with some
extended emission. No CMD information is available for this isolated dwarf,
situated at a distance of $1.32\pm0.7$\,Mpc (3D) from NGC5236.

\subsubsection{ESO269-G037}
Because of stellar crowding in the field of this galaxy it is hard to tell
if the CMD shows young main sequence and upper-AGB stars or if these are
the results of contamination. \citet{karachentsev2002a} classifies this
galaxy as a dSph.

This object was detected in \hi{}  and contains $M_{\rm
HI}=4\times10^5$\,\msol{} in a cloud extending south-east of the optical
centre.  It is the only object of the CenA group where an \hi{} mass below
10$^7$\,\msol{} was detected and it is believed to be presently losing its
\hi{} through ram pressure stripping \citep{bouchard2007}.

We have found some \ha{} emission coinciding with the optical centre of the
dwarf, which is also a high density \hi{} region.  There is also \ha{}
emission in two regions to the West and South-West of the optical center,
surrounding foreground stars. 

After close inspection of the \ha{} image, it seems that the different
seeing conditions for the on-band and the off-band images have created
artifacts around some foreground stars. The emission structure of the West
and South-West features are indeed reminiscent of subtraction difficulties.
We can not rule out this possibility for the central feature. In this case,
however, the emission is slightly extended (FWHM $\sim7^{\prime\prime}$
compared to the FWHM $\sim2.5^{\prime\prime}$ seeing) and does not show
signs of an unsuccessful sky subtraction (\eg{} a flux depression at the
location of the point source). We consider this central emission as real
but confirmation should be sought.

ESO269-G037 is situated at $390\pm250$\,kpc (3D) from NGC4945.

\subsubsection{AM1321-304}
The CMD for this dIm seems to reveal some hints of main sequence and
upper-AGB stars \citep{karachentsev2002a} consistent with the small \hii{}
point source near the center of the dwarf. This galaxy is situated at
$550\pm400$\,kpc (3D) from NGC5236.

\subsubsection{ESO384-G016}
This dS0/Im galaxy has a clearly defined upper-AGB population and some
hints of a young main sequence \citep{karachentsev2007} while the \ha{}
emission is faint and diffuse. \citet{beaulieu2006} found
$6\times10^6$\,\msol{} of \hi{} in this galaxy.  It is situated at $790\pm180$\,kpc
(3D) from NGC5128.

\subsection{Comments on individual objects of the Scl group}

\subsubsection{ESO349-G031}
Only two \ha{} point sources, one centered on the galaxy and one at
$\sim25$\arcsec{} or 390\,pc projected distance (at $D = 3.2$\,Mpc) to the
south-east are detected. The CMD also has clear upper-AGB and main sequence
population \citep{karachentsev2006}. This IBm galaxy is situated $570\pm200$\,kpc
(3D) from NGC300.

\subsubsection{ESO294-G010}
Existing spectroscopic data for this object showed clear detections of the
\ha{} (656.3\,nm) and [O\,{\sc iii}] (500.7\,nm) lines but the signal to
noise ratio did not allow the detection of the second [O\,{\sc iii}]
(495.9\,nm) line \citep[see][Figure 5]{jerjen1998}. These lines reportedly
originate from a feature 18\arcsec{} or 150\,kpc projected distance (at $D
= 1.7$\,Mpc) south of the center of the galaxy. This is in disagreement
with the present results which show \ha{} emission $\sim7$\arcsec{} or
60\,kpc north of the center, more or less coincident with the detected
\hi{} emission \citep{bouchard2005}. Careful inspection of our residual
\ha{} image shows no sign of emission on the southern side.  This dS0/Im
galaxy is situated $130\pm50$\,kpc (3D) from NGC55 and $490\pm140$\,kpc from NGC300.

\subsubsection{DDO6}
There is evidence for diffuse \ha{} emission south-west of the center of
this IB(s)m galaxy and \citet{cote1997} have found $M_{\rm
HI}\,=\,1.2\times10^7$\,\msol{}. The CMD shows a large population of
upper-AGB and young blue main sequence stars \citep{karachentsev2003}. This
object is  $660\pm300$\,kpc (3D) from NGC253.

\subsubsection{ESO540-G032}
There is an unresolved \ha{} source (FWHM $\sim2.4^{\prime\prime}$) at the
centre of this galaxy, for which $M_{\rm HI}=9\times10^5$\,\msol{}
\citep{bouchard2005}. The CMD of this object shows weak evidence for
upper-AGB and blue main sequence populations \citep{jerjen2001a}. It is
situated $640\pm160$\,kpc (3D) from NGC253. 

\subsubsection{AM0106-382}
This isolated galaxy is $2.4\pm1$\,Mpc (3D) from NGC7793. It has strong \ha{}
emission concentrated around four point sources. This dIm galaxy was not
detected by \citet{cote1997} who placed an upper-limit of $M_{\rm
HI}\,<\,1.7\times10^7$\,\msol{} (corrected to a distance of 6.1\,Mpc).
However, the HIPASS survey \citep{barnes2001} reveals a faint \hi{} feature
at $V_\odot\,=\,625$\,\kms{} corresponding to $M_{\rm
HI}\,=\,3\times10^7$\,\msol{} (if at 6.1\,Mpc)\footnote{We use $M_{\rm
HI}\,=\,2.356\times10^6 D^2 S_{\rm HI}$ \citep[c.f.,][]{bouchard2005}.}.

\section{The influence of environment on evolution of dwarf galaxies}
It is a well established fact that the fraction of galaxies sustaining
current star formation decreases towards the projected center of dense
galaxy clusters and this influence can be seen to several virial radii
\citep{ellingson2001, lewis2002, gomez2003, balogh2004a, rines2005}.
Similarly, the morphology-density relation is evident in the Local Group
with the majority of the late-type dwarfs occurring in relatively isolated
locations while the early-type dwarfs are concentrated mainly in the
vicinity of the two large spirals, the Milky Way and M31.  Apart from this
qualitative morphology-density relation, there has otherwise been virtually
no attempt to quantify this relation using local volume galaxies.

\subsection{Physical properties of dwarfs: Quantifying the evolution}
Based on the values presented in Table \ref{objects}, we can derive the
current star formation rate (SFR) for each galaxy.  Here, we follow the
discussion by \citet{skillman2003a}, which, in turn, was based on
\citet{kennicutt1994}:
\begin{equation}
\label{sfr}
SFR~=~\frac{L({\rm H}\,\alpha)}{1.26\times10^{41}\,{\rm ergs}\,{\rm s}^{-1}}\,M_{\odot}\,{\rm yr}^{-1}
\end{equation}
\noindent where SFR is in $M_{\odot}\,$yr$^{-1}$ and $L$(\ha{}), in
erg~s$^{-1}$, is given by:
\begin{equation}
\label{lha}
L({\rm H\,\alpha})~=~4\,\pi\,d^2\,\,F({\rm H\,\alpha})
\end{equation}
\noindent where $d$ is the line of sight distance, in centimeters and the \ha{}
flux, $F({\rm H\,\alpha})$, is in erg~s$^{-1}$\,cm$^{-2}$.

Because SFR is known to correlate with the \hi{} mass, $M_{\rm HI}$, and
blue luminosity, $L_{\rm B}$, one should be careful when interpreting these
parameters. We can, however, define a new set of independent parameters by
taking the ratios of these quantities. These are the galaxy formation
timescale $\tau_{\rm form}$, the gas depletion timescale $\tau_{\rm gas}$
and \hi{} mass to light ratio $M_{\rm HI}/L_{\rm B}$.

Here, 
\begin{equation}
\label{tf}
\tau_{\rm form}~=~\frac{M_{\star}}{\rm SFR}~=~\frac{L_{\rm B}}{\rm SFR}~\left(\frac{M_{\star}}{L_{\rm B}}\right)
\end{equation}
\noindent where $M_{\star}$ is the total stellar mass in solar units
$M_{\odot}$. However, $M_{\star}$ cannot be directly determined. It is
usually inferred from the observable B band luminosity, $L_{\rm B}$, in
solar units and the stellar mass to light ratio $M_{\star}/L_{\rm B}$,
which is a model dependent quantity. To avoid making any assumptions on
this last parameter, we will use
\begin{equation}
\label{tfstar}
\tau_{\rm form}^{\star}~=~\frac{L_{\rm B}}{\rm SFR}~=~\tau_{\rm form}~\left(\frac{M_{\star}}{L_{\rm B}}\right)^{-1}
\end{equation}
\noindent In this context, $\tau_{\rm form}^{\star}$ is the amount of time,
in years, that it would have taken to form the currently observed galaxy if
it always had the current SFR and a stellar mass to light ratio of 1.  We
should bear in mind that any variations seen in $\tau_{\rm form}^{\star}$
can be due to variations in either of $\tau_{\rm form}$ or
$M_{\star}/L_{\rm B}$. For this last parameter, values between 0.1 and 5
are typical for dwarf galaxies \citep{worthey1994}.

Similarly, 
\begin{equation}
\tau_{\rm gas}~=~\frac{M_{\rm HI}}{\rm SFR}
\end{equation}
\noindent is the time, in years, before all the remaining \hi{} gas will be
exhausted if the galaxy retains the current SFR.  We emphasise that both
timescales are instantaneous parameters and by no means represent any past
average values.

\newcommand{\ctd}{log($\rho_L$)}
\newcommand{\pcta}{log~$L$(\ha{})}
\newcommand{\pctb}{SFR}
\newcommand{\pctc}{$M_{HI}$}
\newcommand{\pctd}{$L_{\rm B}$}
\newcommand{\pcte}{log($\tau_{\rm form}^{\star}$)}
\newcommand{\pctf}{log($\tau_{\rm gas}$)}
\newcommand{\pctg}{$M_{HI}/L_{\rm B}$}

\newcommand{\cud}{($L_\odot$ Mpc$^{-3}$)}
\newcommand{\pcua}{(erg s$^{-1}$)}
\newcommand{\pcub}{($10^{-5}$ \msol{} yr$^{-1}$)}
\newcommand{\pcuc}{(10$^6$ \msol)}
\newcommand{\pcud}{(10$^6 L_{\odot}$)}
\newcommand{\pcue}{(yr)}
\newcommand{\pcuf}{(yr)}
\newcommand{\pcug}{(\msol/$L_\odot$)}

\begin{table*}[tbp]
\begin{center}
\caption{Physical properties of CenA group galaxies}\label{physcen}
\begin{tiny}
\begin{center}
\begin{tabular}{l c c c c c c c cc}
\hline
\hline
Galaxy 			& \ctd		& \pcta 		& \pctb 		& \pctc  	     	& \pctd			& \pcte 		& \pctf 		& \pctg			& Ref.\\
       			& \cud		& \pcua			& \pcub 		& \pcuc			& \pcud			& \pcue           	& \pcuf         	& \pcug			&     \\
\hline                                                                                                                                          
{}ESO379-G007 		 &  9.6$\pm$0.1	& 37.4$\pm$0.1	 	& 19$\pm$6		& 33$\pm$6 		& 9.9$\pm$0.8 		& 10.7$\pm$0.2	 	& 11.3$\pm$0.2	 	& 3.4$\pm$0.8	 	& 1 \\
{}ESO321-G014 		 & 10.0$\pm$0.2	& 37.9$\pm$0.1	 	& 65$\pm$20 		& 15$\pm$3 		& 13.3$\pm$1.1 		& 10.3$\pm$0.2	 	& 10.4$\pm$0.2	 	& 1.2$\pm$0.3		& 1  \\
{}ESO381-G018 		 & 10.1$\pm$0.1	& \nodata 		& \nodata 		& \nodata 		& 23$\pm$2 		& \nodata 		& \nodata 		& \nodata 		&   \\
{}ESO381-G020 		 & 10.1$\pm$0.3	& 38.9$\pm$0.1		& 600$\pm$200 		& 22$\pm$30 		& 95$\pm$8 		& 10.2$\pm$0.2	 	& 10.5$\pm$0.1	 	& 2.4$\pm$0.4	 	& 2  \\
{}ESO443-G009 		 & 10.0$\pm$0.5	& \nodata 		& \nodata 		& \nodata 		& 7.9$\pm$0.7 		& \nodata 		& \nodata 		& \nodata 		&   \\
{}ESO219-G010 		 &  9.7$\pm$0.2	& 37.9$\pm$0.1 		& 60$\pm$20 		& $<$0.7 		& 9.8$\pm$1.5 		& 10.2$\pm$0.2	 	& $<$9.1		& $<$0.03 		& 5  \\
{}UGCA319 		 &  9.6$\pm$0.3	& 37.2$\pm$0.2 		& 13.5$\pm$6.0 		& 24$\pm$9 		& 23$\pm$2 		& 11.2$\pm$0.3	 	& 11.3$\pm$0.3 		& 1.1$\pm$0.6	 	& 2  \\
{}DDO161 		 &  9.6$\pm$0.3	& 37.40$\pm$0.15 	& 19.8$\pm$7.0 		& 430$\pm$160 		& 94$\pm$9 		& 11.7$\pm$0.3	 	& 12.3$\pm$0.2	 	& 4.5$\pm$2.3	 	& 2, 4  \\
{}ESO269-G037 		 & 10.9$\pm$0.3	& 36.7$\pm$0.1		& 3.7$\pm$1.2 		& 0.4$\pm$0.1 		& 5.9$\pm$0.5 		& 11.2$\pm$0.2	 	& 10.0$\pm$0.2	 	& 0.07$\pm$0.02 	& 5  \\
{}[CFC97]Cen6 		 &  9.6$\pm$0.2	& \nodata 		& \nodata 		& 42$\pm$5 		& 15.6$\pm$1.4 		& \nodata 		& \nodata 		& 2.7$\pm$0.50	 	& 5  \\
{}NGC4945 		 & 10.6$\pm$0.1	& \nodata 		& \nodata 		& 1100$\pm$200 		& 4500$\pm$900 		& \nodata 		& \nodata 		& 0.24$\pm$0.07 	& 1  \\
{}ESO269-G058 		 & 11.5$\pm$0.1	& \nodata 		& \nodata 		& \nodata 		& 110$\pm$10	 	& \nodata 		& \nodata 		& \nodata 		&   \\
{}CenA-dE1 		 & 10.7$\pm$0.4	& $<$35.7		& $<$0.4 		& $<$0.2		& 2.2$\pm$0.2 		& \nodata 		&\nodata 		& $<$0.07 		& 5  \\
{}ESO269-G066 		 & 11.1$\pm$0.5	& $<$35.7		& $<$0.4 		& $<$0.1 		& 38$\pm$3 		& \nodata 		&\nodata 		& $<$0.003 		& 5  \\
{}HIPASSJ1321-31 	 & 11.6$\pm$0.2	& $<$35.9		& $<$0.6 		& 38$\pm$5 		& 6.2$\pm$0.6 		& $>$12.0 		&$>$12.8		& 6$\pm$1	 	& 1  \\
{}CenA-dE2 		 & 10.3$\pm$0.3	& $<$35.7		& $<$0.5 		& $<$0.6 		& 1.7$\pm$0.3 		& \nodata 		&\nodata 		& $<$0.3 		& 5  \\
{}[KK98]196 		 & 11.4$\pm$0.6	& \nodata 		& \nodata 		& \nodata 		& 8.8$\pm$0.8 		& \nodata 		& \nodata 		& \nodata 		&   \\
{}NGC5102 		 & 10.6$\pm$0.3	& 38.85$\pm$0.08 	& 570$\pm$110 		& 220$\pm$40 		& 1880$\pm$240	 	& 11.5$\pm$0.1	 	& 10.6$\pm$0.1	 	& 0.12$\pm$0.03 	& 1, 6  \\
{}SGC1319.1-4216 	 & 12.5$\pm$1.2	& $<$35.7 		& $<$0.4 		& $<$0.3 		& 13$\pm$2 		& \nodata 		&\nodata 		& $<$0.02 		& 5  \\
{}[KK2000]55 		 & 11.9$\pm$0.9	& \nodata 		& \nodata 		& \nodata 		& 1.0$\pm$0.1 		& \nodata 		& \nodata 		& \nodata		&   \\
{}[CFC97]Cen8 		 & 10.3$\pm$0.3	& $<$35.8 		& $<$0.5 		& $<$0.6 		& 2.7$\pm$0.2 		& \nodata 		&\nodata 		& $<$0.2 		& 5  \\
{}AM1320-230 		 &  9.9$\pm$0.4	& $<$35.8 		& $<$0.5 		& $<$0.5 		& 3.0$\pm$0.2 		& \nodata 		&\nodata 		& $<$0.2 		& 5  \\
{}AM1321-304 		 & 10.7$\pm$0.4	& 36.8$\pm$0.2 		& 4.5$\pm$1.4 		& 20$\pm$10 		& 7.3$\pm$0.7 		& 11.2$\pm$0.2	 	& 11.6$\pm$0.5	 	& 3$\pm$2	 	& 7  \\
{}NGC5128 		 & 10.7$\pm$0.3	& \nodata 		& \nodata 		& 310$\pm$50 		& 16500$\pm$1400 	& \nodata 		& \nodata 		& 0.018$\pm$0.005 	& 1  \\
{}IC4247 		 & 11.6$\pm$1.3	& \nodata 		& \nodata 		& 37$\pm$25 		& 67$\pm$6 		& \nodata 		& \nodata 		& 0.5$\pm$0.4	 	& 7  \\
{}ESO324-G024 		 & 12.4$\pm$0.7	& 36.9$\pm$0.1		& 7.1$\pm$1.6 		& 200$\pm$30	 	& 151$\pm$12	 	& 12.3$\pm$0.1 		& 12.4$\pm$0.1	 	& 1.3$\pm$0.3	 	& 1, 4  \\
{}NGC5206 		 & 10.8$\pm$0.3	& \nodata 		& \nodata 		& $<$5.4 		& 430$\pm$40	 	& \nodata 		& \nodata 		& $<$0.01 		& 2  \\
{}ESO270-G017 		 & 10.3$\pm$0.4	& \nodata 		& \nodata 		& 920$\pm$340 		& 650$\pm$50	 	& \nodata 		& \nodata 		& 1.4$\pm$0.7	 	& 1  \\
{}UGCA365 		 & 13.0$\pm$0.6	& \nodata 		& \nodata 		& 18$\pm$2 		& 27$\pm$2 		& \nodata 		& \nodata 		& 0.66$\pm$0.12 	& 1  \\
{}[KK98]208 		 & 10.9$\pm$0.4	& $<$35.8		& $<$0.5 		& $<$2.1 		& 66$\pm$6 		& \nodata 		&\nodata 		& $<$0.03 		& 5  \\
{}NGC5236 		 & 10.2$\pm$0.4	& 42.23$\pm$0.08 	& 1360000$\pm$200000 	& 10200$\pm$1500 	& 21700$\pm$1800 	& 9.2$\pm$0.1	 	& 8.88$\pm$0.09 	& 0.5$\pm$0.1	 	& 1, 8  \\
{}DEEPJ1337-33 		 & 10.4$\pm$0.3	& \nodata 		& \nodata 		& \nodata 		& 3.9$\pm$0.4 		& \nodata 		& \nodata 		& \nodata 		&   \\
{}ESO444-G084 		 & 10.7$\pm$0.3	& 35.92$\pm$0.09	& 0.7$\pm$0.1 		& 140$\pm$30 		& 33$\pm$3 		& 12.7$\pm$0.1	 	& 13.3$\pm$0.1	 	& 4$\pm$1	 	& 2, 4  \\
{}HIPASSJ1337-39 	 & 10.1$\pm$0.1	& \nodata 		& \nodata 		& \nodata 		& 9.6$\pm$0.9 		& \nodata 		& \nodata 		& \nodata 		&   \\
{}NGC5237 		 & 10.9$\pm$0.5	& \nodata 		& \nodata 		& 21$\pm$3 		& 91$\pm$8 		& \nodata 		& \nodata 		& 0.23$\pm$0.04 	& 2  \\
{}NGC5253 		 & 10.4$\pm$0.1	& 40.13$\pm$0.08	& 22100$\pm$2500	& 140$\pm$20 		& 720$\pm$60 		& 9.51$\pm$0.09 	& 8.80$\pm$0.08 	& 0.19$\pm$0.04 	& 1, 8  \\
{}IC4316 		 & 10.6$\pm$0.3	& \nodata 		& \nodata 		& 36$\pm$6.0		& 32$\pm$6 		& \nodata 		& \nodata 		& 1.1$\pm$0.3	 	& 2  \\
{}NGC5264 		 & 10.5$\pm$0.3	& 36.60$\pm$0.07	& 3.2$\pm$0.5 		& 80$\pm$20 		& 297$\pm$41 		& 13.0$\pm$0.1	 	& 12.4$\pm$0.1	 	& 0.28$\pm$0.08 	& 2, 4  \\
{}[KK2000]57 		 & 11.4$\pm$0.7	& \nodata 		& \nodata	 	& \nodata 		& 1.4$\pm$0.1 		& \nodata 		& \nodata 		& \nodata 		&   \\
{}AM1339-445 		 & 11.5$\pm$0.1	& $<$35.6		& $<$0.3 		& $<$1.2 		& 6.6$\pm$0.6 		& \nodata 		&\nodata 		& $<$0.2 		& 3  \\
{}LEDA166172 		 & 11.4$\pm$0.5	& $<$35.6 		& $<$0.3 		& $<$1.3 		& 0.8$\pm$0.1 		& \nodata 		&\nodata 		& $<$1.5 		& 7  \\
{}ESO325-G011 		 & 10.8$\pm$0.5	& \nodata 		& \nodata 		& 110$\pm$20 		& 45$\pm$4 		& \nodata 		& \nodata 		& 2.4$\pm$0.7	 	& 2  \\
{}CenA-dE3 		 & 10.2$\pm$0.3	& $<$35.8 		& $<$0.5 		& $<$0.5		& 3.3$\pm$0.5 		& \nodata 		&\nodata 		& $<$0.2 		& 5  \\
{}AM1343-452 		 & 11.0$\pm$0.6	& \nodata 		& \nodata 		& $<$0.2 		& 2.3$\pm$0.2 		& \nodata 		& \nodata 		& $<$0.06 		& 5  \\
{}CenA-dE4 		 & 10.7$\pm$1.4	& $<$35.8 		& $<$0.5 		& $<$0.2 		& 2.8$\pm$0.4 		& \nodata 		&\nodata 		& $<$0.06 		& 5  \\
{}CenN 			 & 10.9$\pm$0.2	& \nodata 		& \nodata 		& \nodata 		& 2.1$\pm$1.3 		& \nodata 		& \nodata 		& \nodata 		&   \\
{}HIPASSJ1348-37 	 &  9.8$\pm$0.3	& \nodata 		& \nodata 		& \nodata 		& 9.1$\pm$0.8 		& \nodata 		& \nodata 		& \nodata 		&   \\
{}LEDA166179 		 & 10.8$\pm$0.3	& \nodata 		& \nodata 		& $<$1.5 		& 1.6$\pm$0.1 		& \nodata 		& \nodata 		& $<$1	 		& 7  \\
{}ESO383-G087 		 & 10.8$\pm$0.2	& 37.50$\pm$0.05 	& 25.0$\pm$3.0 		& 104$\pm$16 		& 750$\pm$60	 	& 12.48$\pm$0.09 	& 11.62$\pm$0.08 	& 0.14$\pm$0.03 	& 2, 4  \\
{}HIPASSJ1351-47	 &  8.7$\pm$0.1	& \nodata 		& \nodata 		& \nodata 		& 5.1$\pm$0.5 		& \nodata 		& \nodata 		& \nodata		&   \\
{}ESO384-G016 		 & 10.2$\pm$0.2	& 37.5$\pm$0.1 		& 23$\pm$6 		& 4.4$\pm$0.8 		& 25.6$\pm$1.4 		& 11.0$\pm$0.1	 	& 10.3$\pm$0.1	 	& 0.17$\pm$0.03 	& 5  \\
{}NGC5408 		 &  9.9$\pm$0.2	& \nodata 		& \nodata 		& 320$\pm$50 		& 480$\pm$90	 	& \nodata 		& \nodata 		& 0.7$\pm$0.2	 	& 2  \\
{}UKS1424-460	 	 & 10.1$\pm$0.1	& \nodata 		& \nodata 		& 50$\pm$20 		& 5.0$\pm$0.5 		& \nodata 		& \nodata 		& 10$\pm$4	 	& 2  \\
{}CenA-dE5 		 & 10.0$\pm$0.3	& $<$35.8 		& $<$0.5 		& $<$0.4 		& 1.3$\pm$0.2 		& \nodata 		&\nodata 		& $<$0.3 		& 5  \\
{}ESO222-G010 		 &  9.6$\pm$0.3	& \nodata 		& \nodata 		& 42$\pm$15 		& 9.0$\pm$0.8 		& \nodata 		& \nodata 		& 4.6$\pm$2.4	 	& 2  \\
{}ESO272-G025 		 &  9.7$\pm$0.3	& \nodata 		& \nodata 		& 7$\pm$3 		& 37$\pm$3 		& \nodata 		& \nodata 		& 0.2$\pm$0.1	 	& 5  \\
{}ESO223-G009 		 &  8.8$\pm$0.1	& \nodata		& \nodata 		& 880$\pm$110 		& 78$\pm$7 		& \nodata 		& \nodata 		& 11$\pm$2	 	& 2  \\
{}ESO274-G001 		 &  9.6$\pm$0.1	& \nodata 		& \nodata 		& 330$\pm$50 		& 230$\pm$50	 	& \nodata 		& \nodata 		& 1.4$\pm$0.4	 	& 2  \\
\hline
\end{tabular}
\begin{itemize}
\item {\sc References} --- (1) \citet{koribalski2004}; (2) \citet{cote1997}; (3)
\citet{beaulieu2006}; (4) \citet{lee2003}; 
(5) \citet{bouchard2007};
(6) \citet{macchetto1996}; (7) \citet{banks1999}; (8) \citet{buat2002};
References for $L_{\rm B}$ can be found in Table \ref{astrocen}.
\end{itemize}
\end{center}
\end{tiny}
\end{center}
\end{table*}

\begin{table*}[tbp]
\begin{center}
\caption{Physical properties of Sculptor group galaxies}\label{physscl}
\begin{tiny}
\begin{center}
\begin{tabular}{l c c c c c c c c c}
\hline
\hline
Galaxy 		& \ctd		& \pcta 		& \pctb 		& \pctc        		& \pctd			& \pcte 		& \pctf 		& \pctg			& Ref.\\
       		& \cud		& \pcua       		& \pcub			& \pcuc 		& \pcud			& \pcue                 & \pcuf                 & \pcug			&     \\
\hline                                                                                                                  
{}WHIB2317-32 	& 10.3$\pm$0.4	& $<$35.4 		& $<$0.2 		& 33$\pm$21 		& 1.4$\pm$0.1 		& $>$11.8 		&$>$13.2 		& 24$\pm$21	 	& 1  \\
{}UGCA438 	& 10.0$\pm$0.1	& \nodata 		& \nodata 		& \nodata 		& 22.5$\pm$2 		& \nodata 		& \nodata 		& \nodata		&   \\
{}ESO347-G017 	&  8.8$\pm$0.2	& 38.89$\pm$0.08	& 610$\pm$110 		& 120$\pm$20 		& 103$\pm$10	 	& 10.2$\pm$0.1	 	& 10.3$\pm$0.1	 	& 1.2$\pm$0.3 		& 1, 2  \\
{}UGCA442 	& 10.3$\pm$0.3	& 37.14$\pm$0.08 	& 11.0$\pm$2.0 		& 230$\pm$40 		& 105$\pm$10	 	& 12.0$\pm$0.1	 	& 12.3$\pm$0.1	 	& 2.2$\pm$0.6 		& 1, 3  \\
{}ESO348-G009 	&  8.8$\pm$0.1	& 37.54$\pm$0.09 	& 28$\pm$5 		& 84$\pm$13 		& 240$\pm$20	 	& 12.0$\pm$0.1	 	& 11.5$\pm$0.1 		& 0.34$\pm$0.08		& 1, 2  \\
{}NGC7793 	& 10.2$\pm$0.1	& 38.6$\pm$0.1	 	& 1.5$\pm$0.2 		& 1000$\pm$200 		& 3100$\pm$300	 	& 14.3$\pm$0.1	 	& 13.8$\pm$0.1 		& 0.32$\pm$0.08		& 4, 5  \\
{}[CFC97]SC18	& 10.6$\pm$0.2	& 36.46$\pm$0.07 	& 2.3$\pm$0.4 		& 5.0$\pm$1.0 		& 0.9$\pm$0.1 		& 10.6$\pm$0.1	 	& 11.3$\pm$0.1 		& 5.7$\pm$1.5 		& 1, 2  \\
{}ESO349-G031 	&  9.8$\pm$0.2	& 36.7$\pm$0.1 		& 3.8$\pm$1.2 		& 4.4$\pm$1.2 		& 7.0$\pm$0.6 		& 11.3$\pm$0.2 		& 11.1$\pm$0.2 		& 0.6$\pm$0.2 		&  1 \\
{}NGC24 	&  8.9$\pm$0.3	& 39.3$\pm$0.3  	& 1800$\pm$1100 	& 590$\pm$50 		& 820$\pm$80	 	& 10.1$\pm$0.5 		& 10.5$\pm$0.3		& 0.7$\pm$0.2 		& 4, 6  \\
{}NGC45 	&  9.3$\pm$0.2	& \nodata 		& \nodata 		& 1520$\pm$200 		& 1320$\pm$120	 	& \nodata		& \nodata		& 1.2$\pm$0.3 		& 4  \\
{}NGC55 	& 10.2$\pm$0.3	& 40.44$\pm$0.06 	& 21800$\pm$3000 	& 1500$\pm$300 		& 1670$\pm$150	 	& 9.8$\pm$0.1 		& 9.8$\pm$0.1 		& 1.0$\pm$0.2 		& 4, 8  \\
{}NGC59 	& 10.6$\pm$0.3	& 38.57$\pm$0.09 	& 300$\pm$60 		& 17$\pm$4 		& 198$\pm$6 		& 10.8$\pm$0.1 		& 9.8$\pm$0.1 		& 0.08$\pm$0.02		& 2, 9  \\
{}ESO410-G005 	& 10.6$\pm$0.1	& $<$34.8 		& $<$0.1 		& 0.7$\pm$0.1 		& 6.8$\pm$0.6 		& $>$13.1		&$>$12.1		& 0.11$\pm$0.02		& 10  \\
{}Scl-dE1 	& 11.1$\pm$0.5  & $<$35.5 		& $<$0.3 		& $<$0.19 		& 2.3$\pm$0.4 		& \nodata		&\nodata 		& $<$0.08		& 10  \\
{}ESO294-G010 	& 10.7$\pm$0.2	& 36.7$\pm$0.1  	& 3.7$\pm$1.0 		& 0.30$\pm$0.03 	& 2.8$\pm$0.1 		& 10.9$\pm$0.1 		& 9.8$\pm$0.1 		& 0.08$\pm$0.01		& 10  \\
{}ESO473-G024 	& 10.7$\pm$0.5	& 38.20$\pm$0.08 	& 126$\pm$22 		& 64$\pm$10 		& 23$\pm$2 		& 10.3$\pm$0.1 		& 10.7$\pm$0.1 		& 2.8$\pm$0.7 		& 1, 2  \\
{}[CFC97]SC24 	& 10.3$\pm$0.2	& 35.42$\pm$0.08 	& 0.2$\pm$0.05 		& 7.7$\pm$1.2 		& 0.3$\pm$0.1 		& 11.1$\pm$0.1 		& 12.6$\pm$0.1 		& 27$\pm$6 		& 1, 2  \\
{}IC1574 	&  9.8$\pm$0.3  & 38.3$\pm$0.1 		& 160$\pm$35 		& 42$\pm$8 		& 70$\pm$6 		& 10.6$\pm$0.2 		& 10.4$\pm$0.1		& 0.6$\pm$0.2 		& 1, 11  \\
{}NGC247 	& 11.0$\pm$0.4	& \nodata 		& \nodata 		& 2400$\pm$400 		& 3000$\pm$300 		& \nodata		& \nodata		& 0.8$\pm$0.2 		& 4  \\
{}NGC253 	& 10.3$\pm$0.3	& \nodata 		& \nodata 		& 2500$\pm$400 		& 16700$\pm$1500 	& \nodata		& \nodata		& 0.15$\pm$0.04		& 4  \\
{}ESO540-G030 	&  9.9$\pm$0.3	& $<$35.3 		& $<$0.1 		& 0.8$\pm$0.1 		& 4.6$\pm$0.3 		& $>$12.5		&$>$11.7		& 0.17$\pm$0.02		& 10  \\
{}DDO6 		& 10.2$\pm$0.4	& 36.9$\pm$0.1	 	& 6.6$\pm$1.9 		& 11.8$\pm$2.0 		& 14.8$\pm$1.4 		& 11.4$\pm$0.2 		& 11.3$\pm$0.2 		& 0.8$\pm$0.2 		& 1  \\
{}ESO540-G032 	& 10.8$\pm$0.3  & 36.0$\pm$0.2	 	& 0.8$\pm$0.3 		& 0.9$\pm$0.1 		& 5.8$\pm$0.4 		& 11.8$\pm$0.2 		& 11.1$\pm$0.2 		& 0.19$\pm$0.03		& 10  \\
{}NGC300 	& 10.1$\pm$0.4	& \nodata 		& \nodata 		& 2200$\pm$400 		& 1900$\pm$200 		& \nodata 		& \nodata		& 1.1$\pm$0.3 		& 4  \\
{}AM0106-382 	&  8.8$\pm$0.3	& 38.5$\pm$0.1 		& 250$\pm$80 		& 30$\pm$6 		& 19$\pm$2 		& 9.9$\pm$0.2 		& 10.1$\pm$0.2 		& 1.6$\pm$0.4 		& 1  \\
{}NGC625 	&  9.6$\pm$0.1	& 39.78$\pm$0.07 	& 4800$\pm$750 		& 118$\pm$17 		& 570$\pm$50	 	& 10.1$\pm$0.1 		& 9.4$\pm$0.1		& 0.21$\pm$0.05 	& 1, 2  \\
{}ESO245-G005 	&  9.4$\pm$0.2	& 39.06$\pm$0.08 	& 920$\pm$160 		& 400$\pm$60 		& 260$\pm$20 		& 10.5$\pm$0.1 		& 10.6$\pm$0.1		& 1.6$\pm$0.4		& 1, 12  \\
\hline
\end{tabular}
\begin{itemize}
\item {\sc References} --- (1) \citet{cote1997}; (2) \citet{skillman2003a}; (3)
\citet{lee2003}; (4) \citet{koribalski2004}; (5) \citet{buat2002}; (6)
\citet{romanishin1990}; (7) \citet{chemin2006}; (8) \citet{miller2003}; (9) \citet{beaulieu2006};
(10) \citet{bouchard2005};
(11) \citet{vanzee2000}; (12) \citet{miller1996}; References for $L_{\rm B}$ can be found in Table \ref{astroscl}.
\end{itemize}
\end{center}
\end{tiny}
\end{center}
\end{table*}

All the above mentioned parameters, namely $M_{\rm HI}$, $L({\rm
H\,\alpha})$, SFR, $\tau_{\rm gas}$ and $\tau_{\rm form}^{\star}$, are
listed in Table \ref{physcen} for galaxies of the CenA group and in Table
\ref{physscl} for the members of the Scl group.

\subsection{Luminosity density: Quantifying the environment}
Environment is a difficult concept to quantify and the methods used to
describe it vary greatly.  In distant clusters most authors use either
projected distance from the cluster center or the local surface number
density \citep[\eg{}][]{rines2005}.  The first assumes a spherically
symmetric distribution.  The second assumes that the influence from each
galaxy is strictly the same and, to minimize the impact of this assumption,
authors generally apply a cutoff by only considering bright galaxies.

In the present study, we can not afford making either of these assumptions.
There are two main reasons: first, these groups are not symmetric
\citep[see][]{jerjen1998}. Second, there is only a handful of large
galaxies in the groups and discarding dwarfs from the sample would result
in considering only one or two objects per group.  Fortunately, independent
radial distances measurements are available for most galaxies in the local
volume (see references in Tables \ref{astrocen} and \ref{astroscl}). Using
this information, we can estimate the 3D local galaxy density within the
groups.  Since our aim is to compare galaxies from very different
environments, it is important to define a parameter that is truly
representative of the local surroundings of each studied galaxy.

The local luminosity density $\rho_{L}$ (in $L_{\odot}\,{\rm Mpc}^{-3}$) at
the position of galaxy $i$ as a function of the B band luminosity ($L_B$,
in $L_{\odot}$) and 3D distance ($R$, in Mpc) is defined by:
\begin{equation}
\rho_{L}(i)~=~\sum_{j\,\ne\,i} \frac{L_B(j)}{(4/3)\,\pi\,R_{ij}^3}
\label{density}
\end{equation}
\noindent where the sum goes over all neighbour galaxies $j$.  This
quantity accurately and systematically describes the environment of each
galaxy in the groups.  Here, $R_{ij}$ is simply defined as the spatial
separation of galaxies $i$ and $j$:
\begin{equation}
R_{ij}~=~ \sqrt{D_i^2 + D_j^2 - 2 D_i\,D_j\,{\rm Cos}(\theta_{ij})}
\label{radius}
\end{equation}
\noindent with $\theta_{ij}$ being the angular separation between a pair of
galaxies. Unfortunately, not all values of $R_{ij}$ can be calculated
because not all radial distances $D$ have been measured. In these cases, we
adopted $D\,=\,4.4\pm0.8$ Mpc for galaxies of the CenA group and
$D\,=\,3.7\pm1.6$ Mpc for galaxies of the Scl group. These correspond to
the mean distance to all galaxies in the groups with known distances and an
interval which encompasses $\sim66\%$ of those distances.

The value of $\rho_{L}$ varies by some 5 orders of magnitudes. It ranges
from 10$^8\,$L$_{\odot}\,{\rm Mpc}^{-3}$ for the most isolated objects in
the groups, such as the Scl dwarf ESO348-G009, and goes up to
10$^{13}\,$L$_{\odot}\,{\rm Mpc}^{-3}$ for NGC5264, which is at roughly
75\,kpc (true distance) from NGC5236 (M83, $M_{\rm B}=-20$).  As such, we
are not probing any regime of extremely high densities.  For comparison,
the Sagittarius dwarf is at $\sim25$\,kpc \citep{monaco2004} from the Milky
Way ($M_{\rm B}=-20.6$), which gives $\rho_{L}\sim10^{15}$ and the Small
Magellanic Cloud, at $\sim60$\,kpc, has $\rho_{L}\sim10^{14}$.  Moreover,
the clusters studied by \citet{rines2005} showed cases where galaxy surface
number densities reached well above 100 Mpc$^{-2}$ by only considering
galaxies with $M_{\rm K}\le-22.7$. This is roughly equivalent to having the
closest neighbour at 100 kpc projected distance and would correspond to
$\rho_{L}[K]\gtrsim10^{13}$.  The present work is therefore extending
previous studies to regimes of lower densities that have not been tested
before.

\begin{figure}[tbp]
\begin{center}
\includegraphics[width=0.45\textwidth]{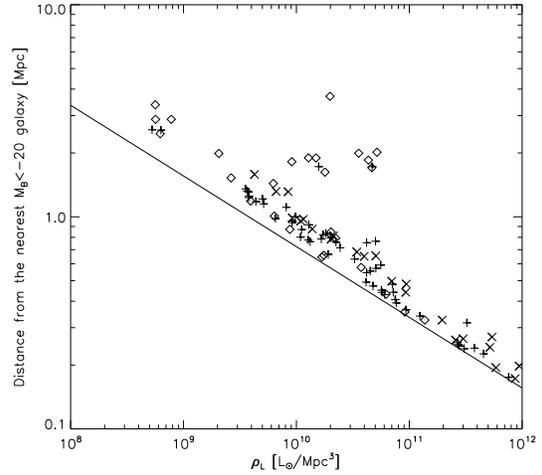}
\caption{The 3D distance to the nearest bright ($M_{\rm B}\,<\,-20$) galaxy compared to $\rho_{L}$ for all galaxies in the Local (\emph{pluses}), CenA (\emph{crosses}) and Scl (\emph{diamonds}) groups. The diagonal line represents the contribution of a single $M_{\rm B} = -20$ galaxy.}\label{dist_vs_den}
\end{center}
\end{figure}

Figure \ref{dist_vs_den} presents a comparison between $\rho_{L}$ and the 3D distance to the nearest $M_{\rm B} < -20$ galaxy. 
Within our target galaxy groups, the environment of objects situated less than ~350\,kpc from a bright massive object ($\rho_{L} > 10^{11}\,$L$_{\odot}\,{\rm Mpc}^{-3}$) is equally well describe by either the 3D distance to that galaxy or by $\rho_{L}$. In these cases, the bright galaxy clearly dominates $\rho_{L}$ and the contribution from the rest of the group is negligible. However, for objects further than 350\,kpc there may be significant discrepancies between $\rho_{L}$ and the 3D distance to the nearest $M_{\rm B} < -20$ galaxy. This is particularly true for Sculptor galaxies where no clearly dominant object may be defined as the group center and where the nearest massive galaxy may be situated in another group, a situation we have not considered.
Presumably, this effect may become more important as we try to include galaxies from even lower density environment.

Strictly speaking, these values of $\rho_{L}$ are lower limits because we
have an incomplete list of group members. The missing galaxies, those that
are yet to be discovered, will, however, be of low luminosity and therefore
have minimal impact on $\rho_{L}$.

\subsection{Physical properties as a function of environment}

\begin{figure}[tbh]
\begin{center}
\includegraphics[angle=90, width=0.5\textwidth]{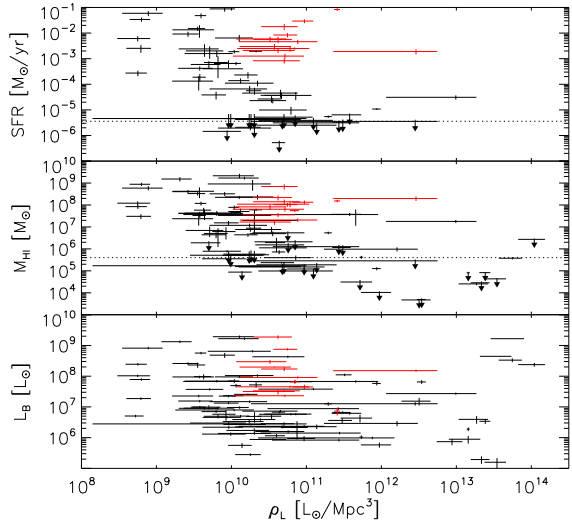}
\caption{The current star formation rate (\emph{top}), \hi{} mass
(\emph{center}) and B-band luminosity (\emph{bottom}) for galaxies of the
Centaurus A, Sculptor and Local Group, as a function of $\rho_L$. 
For the top two panels, the \emph{grey dotted horizontal} line shows the typical detection limit, see text for details. Some galaxies were highlighted (\emph{red markers}) as these objects may be on their first infall onto their respective group.}\label{sfrvsden}
\end{center}
\end{figure}

\begin{figure}[tbh]
\begin{center}
\includegraphics[angle=90, width=0.5\textwidth]{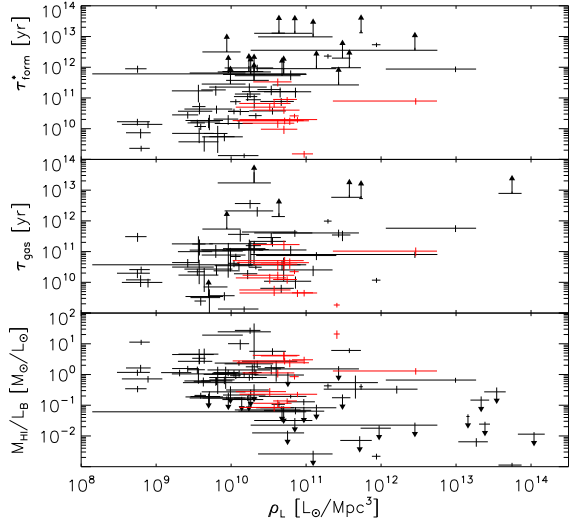}
\caption{The galaxy formation timescale (\emph{top}), gas depletion
timescale (\emph{center}) and \hi{} mass to light ratios (\emph{bottom})
for galaxies of the Centaurus A, Sculptor and Local Group, as a function of
$\rho_L$. 
Some highlighted galaxies (\emph{red markers}) may be on their first infall onto their respective group.}\label{ttvsden}
\end{center}
\end{figure}

The values of SFR, $M_{\rm HI}$ and $L_{\rm B}$ are plotted against
$\rho_{L}$ in Figure \ref{sfrvsden}. The same is done for $\tau_{\rm
form}^{\star}$, $\tau_{\rm gas}$ and $M_{\rm HI}/L_{\rm B}$ in Figure
\ref{ttvsden}. The figures include data from the CenA and Scl group (Tables
\ref{physcen} and \ref{physscl}) and, to increase the significance of the
results, from the Local Group \citep[][ and references therein]{mateo1998}.

The dotted horizontal lines in the first 2 panels of Figure \ref{sfrvsden} represent the typical detection limit for SFR and $M_{\rm HI}$. 
On average, the lowest detectable \ha{} flux in our sample is 
$F({\rm H\,\alpha})\sim2\times10^{-16}$\,ergs~s$^{-1}$\,cm$^{-2}$ (Table
\ref{objects}) and the average distance of these galaxies is
4.13\,Mpc (Tables \ref{astrocen} and \ref{astroscl}). This means that the
typical detection limit is SFR$=3.6\times10^{-6}$\,\msol\,yr$^{-1}$. The
limit for \hi{} flux is considered to be $\sim$100\,mJy\,\kms{}
\citep{bouchard2005, bouchard2007}, which brings the typical \hi{} mass
detection limit to $M_{\rm HI}=4.0\times10^5$\,\msol.

These figures show a very large scatter in the data points and no directly obvious
correlation can be seen. We recognise that this scatter is largely due to
intrinsic variations from source to source and that the factors are not
solely a function of environment.  It is also important to say that
$\rho_{L}$ is a static description of the present environment and contains
no historical information on its evolution. Nevertheless, this is the best that
can be done short of a full orbit reconstruction for each objects in the
groups \citep[\eg][]{peebles1990}. Yet, however uncertain this
environmental parameter may be, if the quantities in Figures \ref{sfrvsden}
or \ref{ttvsden} have any $\rho_{L}$ dependence, then some level of
correlation is expected.

\begin{table*}[tbp]
\begin{center}
\caption{Correlation coefficients. See text for details.}
\label{fits}
\begin{tabular}{l r @{~~} l r @{~~} l r @{~~} l}
\hline
\hline
					& \multicolumn{2}{c}{Excl. ND} & \multicolumn{2}{c}{Incl. ND} & \multicolumn{2}{c}{`Genuine' ND}\\
					&  $R_s$ & ($p$) &  $R_s$ & ($p$) &  $R_s$ & ($p$) \\
\hline
SFR					& $-0.24$ & (0.075)	&	$-0.39$ & (0.0003)			&	$-0.38$ & (0.001)			\\
$M_{\rm HI}$			& $-0.38$ & (0.002)	&	$-0.57$ & ($<10^{-5}$)		&	$-0.50$ & ($<10^{-5}$)		\\
$L_{\rm B}$			& $-0.24$ & (0.01)	& 	 \multicolumn{2}{c}{\nodata}	& 	 \multicolumn{2}{c}{\nodata}	\\
$\tau_{\rm form}^{\star}$	& $0.30$ & (0.03)	&	$0.46$ & (0.0002)			&	$0.43$ & ($0.0001$)		\\
$\tau_{\rm gas}$		& $0.13$ & (0.31)	&	$0.22$ & (0.05)				&	$0.17$ & (0.14)				\\
$M_{\rm HI}/L_{\rm B}$	& $-0.25$ & (0.04)	&	$-0.50$ & ($<10^{-5}$)		&	$-0.45$ & ($<10^{-5}$)		\\
\hline
\end{tabular}
\end{center}
\end{table*}

To investigate this possibility, we have used a Spearman Rank correlation test. This assesses whether a monotonic function could accurately describe the data without making any assumptions on what the exact relation might be. A correlation coefficient ($R_s$) with a value approaching $R_s \sim \pm1$ means that there is a strong correlation (or anti-correlation) between the two given quantities, while $R_s \sim 0$ implies that the quantities are independent.  
The results of these correlation measurements can be found in Table \ref{fits}, along with the probability $p$ that this corresponds to the null hypothesis (no correlation) using a random permutation test.  
In order to determine the effects of the non-detections on $R_s$, the analysis was done in three different ways: Firstly by ignoring any non-detections (col. 1 of Table \ref{fits}), secondly by including the non-detections at the value of their detection limit (col. 2) and thirdly by considering them as genuine non-detections (SFR\,=\,0\,\msol\,yr$^{-1}$ or $M_{\rm HI}=0$\,\msol, col. 3).

As the detection limits are not spread evenly in $\rho_{L}$ but are
slightly shifted to the right hand side of the plots, the correlations are all
significantly better when these limits are included in the correlation measurements.  Since $L_{\rm
B}$ is known for each system, the distinction between the various methods is
irrelevant for this quantity.

Regardless of the method used to handle detection limits, our data show that there are strong levels of correlation between $\rho_L$ and SFR, $L_{\rm B}$, $\tau_{\rm form}^{\star}$ and $M_{\rm HI}/L_{\rm B}$: the low corresponding $p$ values rule out any possibility of random occurrence. For SFR, the correlation significance may seem weak ($p\,\sim\,0.075$) when detection limits are not considered but the $p$ value substantially decreases when these limits are included.  Only for $\tau_{\rm gas}$ does the data not indicate any significant correlation. 

Although highly significant, the values of $R_s$ remain on the low side ($0.2 < | R_s | < 0.6$) and do not approach 1. This is caused by the scatter in the data and may be a quantitative way of saying that other, probably source-specific parameters (\ie{} not linked to environment) have not been considered.\\ 

On a more specific note, the dependance of SFR and $M_{\rm HI}$ on $\rho_L$ ($R_s \sim -0.4$ and $R_s \sim -0.5$ respectively) are coherent with previous findings: galaxies in high density regions harbour less current star formation \citep{rines2005} and less \hi{} \citep{gavazzi2006} than their lower density counterparts. 
This is also consistent with the morphology-density relation,
where early-type dwarfs are more concentrated near larger members of groups
than late-type dwarfs. Indeed, early-types are known to have low current star formation, low \hi{} mass and overall old stellar population. Finally, the anti-correlation of $L_{\rm B}$ and $\rho_{L}$ ($R_s \sim -0.2$) indicates a slight tendency of having brighter objects in the outskirts of the groups. 
Assuming that this is not simply driven by a selection effect --- given the area of sky to be surveyed brighter galaxies are easier to find in group outskirts --- 
we note that, for a given baryonic
mass, galaxies with a dominant young stellar population have higher values
of $L_{\rm B}$ than older ones, because young stars contribute strongly to
$L_{\rm B}$.  Therefore, knowing that galaxies with higher star formation
rates tend to be in the outskirts of groups, the dependance of $L_{\rm B}$
on $\rho_{L}$ should not blindly be interpreted as a variation of baryonic 
mass with density. Ideally, such a relation should be measured using the near
infrared (\eg{} H band), as $L_{\rm H}$ is a much better tracer of stellar
mass than $L_{\rm B}$ \citep{kirby2008}.  In a more general context, these
results are consistent with other recent evidence that early-type giant
galaxies in high density environments are both older \citep{thomas2005,
clemens2006} and fainter \citep{bernardi2006} than their field
counterparts. 
Overall, this is at odds with $\Lambda$CDM hierarchical merger simulations which predicted an increase of dark matter halo masses (and presumably baryonic luminosity) with increasing density \citep{lemson1999,maulbetsch2007}.

From the lower panel of Figure \ref{ttvsden}, we see that galaxies at low values of
$\rho_{L}$ have higher gas mass fraction then the ones with higher $\rho_{L}$.
There are three possible explanations for this. First, the gas may have
been stripped from the dwarfs by various mechanisms. Such stripping process
are likely to be more prevalent at high $\rho_{L}$. Second, the fraction
of gas that is ionised may be greater for galaxies in the central regions
of groups as the ambient ionising flux level might be higher. Finally, the
gas may have been used up more efficiently in early star formation. We note
that the quantity $\rho_{L}$ should, in principle, correlate
with various physical factors.  The density of the intergalactic medium (IGM),
responsible for ram pressure, will presumably increase with increasing
$\rho_{L}$.  Similarly, the tidal fields, which are due to variations in
the gravitational field, have the same $R^{-3}$ dependence as in
Equation \ref{density}. Additionally, $\rho_{L}$ is also a measure of the
radiation received from the neighbouring galaxies.  Hence, 
environments with large $\rho_{L}$ values provide the mechanisms needed for both removing and ionising
the ISM in the dwarfs (assuming a relation between B band and ionising
flux). The third possibility, the gas being transformed more efficiently in
early star formation epochs, would make the dwarfs in high $\rho_{L}$ regions have
greater stellar masses for the same $L_{\rm B}$ than those in the outskirts
of groups.  Again, the relation between $L_{\rm H}$ and $\rho_{L}$ would
allow investigation of this issue by permitting direct estimates of the
baryonic masses.

Interestingly, $\tau_{\rm form}^{\star}$ is comparable to or larger than
the age of the Universe for almost all sample galaxies. This is a sign that
the current SFR is lower than the past average rate in most galaxies.
Similarly, $\tau_{\rm gas}$ is also comparable to or greater than the age
of the Universe.  However, as $\tau_{\rm form}^{\star}$ increases with
$\rho_{L}$, $\tau_{\rm gas}$ remains more or less constant, \ie{} $\rho_{L}$ does
not affect the rate at which gas is being used up by star formation in
present day dwarf galaxies. This could have also been seen from the fact
that $M_{\rm HI}$ and SFR have similar $\rho_{L}$ dependencies.
Unsurprisingly, we also note that $M_{\rm HI}/L_{\rm B}$ ratio decreases.
So although the star formation seems to be lower at high $\rho_{L}$,
the neutral gas fraction (as opposed to absolute quantity) available for
further star formation is lower than average. This supports the idea that
the \hi{} must have been depleted in an early and strong burst of star
formation or is being removed by external mechanisms.\\

In the upper panel of Figure \ref{sfrvsden} there appears to be a sub-sample of galaxies, identified by red markers, that have higher SFR values than might be expected for their $\rho_{L}$ values. In particular, 
there are 14 galaxies with SFR $> 10^{-3.5}$\,\msol{}\,yr$^{-1}$ and $\rho_L > 10^{10.5}$ (Figure \ref{sfrvsden}). These objects do not belong to any given group: IC10, IC1613, NGC3109 and NGC6822 are in the Local Group; NGC59 and ESO473-G024 are in the Sculptor group; and ESO324-G024, ESO444-G084, NGC5237, NGC5264, IC4316, ESO325-G011, ESO383-G087, NGC5102 belong the CenA group. These galaxies also have higher \hi{} mass than the overall sample average. 
This could be explained if they come from regions of lower $\rho_L$ than most of their counterparts, either because of highly elliptical orbits or they are on their first infall onto the group. One would expect such galaxies to have colder ISM, since they would have been overall less affected by ISM warming mechanisms, \eg{} tidal stirring and neighbouring galaxy radiation. On top of being a prerequisite for star formation, cold ISM should  also be more resilient to gas removal by ram pressure since its density would be higher \citep[see][]{gunn1972}.

Moreover, if these 14 galaxies that are potentially on their first group infall are excluded, then 
$\tau_{\rm gas}$ is no longer largely independent of $\rho_L$ (cf.\ Table \ref{fits}) but instead rises significantly with increasing $\rho_L$ ($R_s\,=\,0.36, p\,=\,0.009$). 
Whether this is driven by the low SFR or the lack of ISM at high $\rho_L$ is unclear.

\citet{grebel2003} also noted that the removal of the ISM from a given
dwarf would not produce a non-rotating object and that therefore the
late-type and early-type dwarfs should be considered as intrinsically
different objects. However, as previously noted, the tidal fields should
also correlate with $\rho_{L}$. These have been known to remove angular
momentum from low total mass objects \citep{read2005}. We would therefore expect
a dependence of galaxy kinematics on nearby galaxy density. This can be measured
observationally: given a sufficient sample, suitably normalised measures of
rotation velocity should decrease with increasing $\rho_{L}$.

\subsection{Centaurus A vs Sculptor: An environmental comparison}
As it can be implied from Figure \ref{dist}, the scarce population of the
Scl group makes for a very different environment to the denser distribution
of the CenA group. The detailed galaxy phase-space distributions (Tables
\ref{astrocen} and \ref{astroscl}) reveals the extent of the difference. On
the one hand, Scl is a loose filament of galaxies \citep{jerjen1998} that
more or less obeys the Hubble flow \citep[$H =
75\,$\kms\,Mpc$^{-1}$]{karachentsev2003}. On the other hand, the CenA group
is a similar environment to that of the Local Group.  CenA has a slightly
higher total mass then the Local Group \citep{karachentsev2002a}, but an
otherwise similar two central body structure.\\

\begin{figure}[tb]
\begin{center}
\includegraphics[angle=90, width=0.5\textwidth]{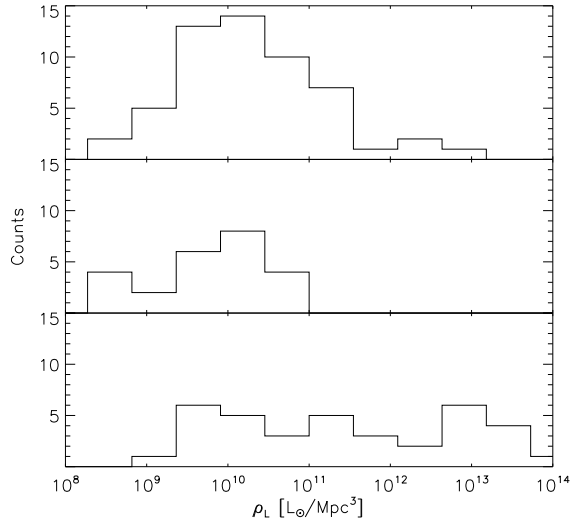}
\caption[Density comparison for the CenA, Scl and Local Group]{Comparison
of the $\rho_{L}$ distribution for the CenA (\emph{top}), Scl (\emph{center}) and Local
Group (\emph{bottom}).}
\label{den}
\end{center}
\end{figure}

Figure \ref{den} shows a breakup of the $\rho_{L}$ distribution for each
group. It can be seen that the method devised in this paper has the merit
of not assuming that all galaxies in each group are in the same situation.
Indeed, analysing the environment on a `per galaxy' basis shows that both
groups have regions with equally low $\rho_{L}$ values, yet, as expected, the mean
$\rho_{L}$ of the Scl group is lower than that of CenA. This latter group,
however, lacks the high $\rho_{L}$ objects seen in the Local Group. These
missing galaxies are, presumably, optically faint dSphs lying close to NGC5128
(CenA) and NGC5236 (M83). Their low surface brightness and relatively large
angular size makes them very hard to detect and thorough searches are
needed to better constrain the high end of the density distribution in the
CenA group.  Note that the addition of low surface brightness galaxies is
unlikely to have any notable effects on the currently measured $\rho_{L}$
distribution.\\

It should also be noted that the low end of the $\rho_{L}$ spectrum is a
matter of definition. No `field' galaxies were included in this study and,
therefore, the cutoff at $10^8 - 10^9\,L_\odot{}\,$Mpc$^{-3}$ is an
artificially imposed one. Nevertheless, the effects of environment can be
seen even at these lowest $\rho_{L}$ values, \ie{} no clear departure from the presumed correlations can be observed in Figures \ref{sfrvsden} and \ref{ttvsden}. One would have expected a turnoff
at low $\rho_{L}$ where the effects of environment become negligible.
The correlations seen in Figure \ref{ttvsden} suggest that environment
accelerates the evolution of dwarfs: galaxies encountered in high $\rho_{L}$ regions
have already formed most of their stars and are transforming their remaining \hi{}
reserve inefficiently at a very slow rate. This implies that the global effect a galaxy
has on its surrounding environment extends very far, at least as far as
1\,Mpc for a $10^9\,L_{\odot}$ galaxy.  Indeed, the lack of a change of
slope at low $\rho_{L}$ in the relation of Figures \ref{sfrvsden} and
\ref{ttvsden} indicates we have not probed $\rho_{L}$ regimes where dwarf galaxy
evolution is not influenced by environment.\\

\section{Conclusions}
The results of \ha{} imaging of Scl and CenA dwarf galaxies were presented and the data were analysed in the context of the morphology-density relation.
The main results are:
\begin{enumerate}
\item{Of the 30 observed objects we have detected \ha{} emission in 13.
These consist of five late-type dwarfs in the CenA group, three in the Scl
group, two early-type dwarfs in CenA, one mixed-type in CenA and two in
Scl.}
\item{The density of the surrounding environment (\ie{} the luminosity and
proximity of other nearby galaxies, as measured by $\rho_{L}$) significantly influences the properties of dwarfs.
Dwarf galaxies in high $\rho_{L}$ regions of the CenA and Scl group  have, in
general, lower star formation rates, lower \hi{} masses, lower \hi{} mass
to luminosity ratios, as well as higher star formation timescale.  The gas
depletion timescale is unaffected by $\rho_{L}$.  These effects
can even be observed out to the lowest probed densities
($\rho_{L}\,\sim\,10^9\,L_{\odot}$\,Mpc$^{-3}$).}
\item{While there are clearly a number of unaccounted parameters that
contribute to the scatter in the factors determining galaxy morphology, an
environmental dependence is evident.  One of the unaccounted factors is
likely to be the orbital history of each object: we only see a snapshot but
orbit averaged values of quantities such as $\rho_{L}$ are really what is
required. }
\item{
Some galaxies do not follow the relation between instantaneous star formation rate and $\rho_{L}$ that would be expected by taking the morphology-density relation at face value.
These galaxies also have larger \hi{} masses than the sample average and may be objects on their first infall onto the galaxy groups: some of their intrinsic properties (\eg{} interstellar medium temperature) may be a reflection of this.} 
\item{The centrally located dwarfs have lower current star formation rates.
Consequently, for the same total stellar mass the objects at high $\rho_{L}$
probably formed most of their stars at earlier times than the ones in low
$\rho_{L}$ regions.}
\item{Our data indicate that there is a weak anti-correlation between B band luminosity and $\rho_{L}$. This may be at odds with $\Lambda$CDM which predicts and increase in total galaxy mass with increasing density. H band photometry of these objects is needed to further investigate this possibility.}
\end{enumerate}

\acknowledgments
We would like to thank St\'ephanie C\^ot\'e for the use of unpublished data, Mike Dopita, Erwin de Blok, Eduard
Westra and Philippe Prugniel for helpful discussions as well as the referee for the highly valuable suggestions, leading to a significant improvement in the paper. This research
has been supported in part by the Australian Research Council through
Discovery Project grant DP0343156.

\bibliographystyle{apj}

\end{document}